%% file: core.tex
\documentclass[journal]{IEEEtran}

\usepackage[T1]{fontenc}% optional T1 font encoding

\usepackage{cite}
\usepackage[pdftex]{graphicx}
\usepackage{amsmath,amssymb,amsfonts}
\usepackage{array} 
\usepackage{epstopdf}
\usepackage{multirow}
\usepackage{subfiles}

\input{sections/commands}

\ifCLASSOPTIONcompsoc
\usepackage[caption=false,font=normalsize,labelfon
t=sf,textfont=sf]{subfig}
\else
\usepackage[caption=false,font=footnotesize]{subfi
g}
\fi

\begin{document}

\title{Sinogram Enhancement with Generative Adversarial Networks using Shape Priors}

\author{Emilien~Valat,
        Katayoun~Farrahi,
        and~Thomas~Blumensath% <-this % stops a space
\thanks{Emilien V., Katayoun F. and Thomas B. are with the University of Southampton, Southampton, United Kingdom}
\thanks{This work is supported by the DSTL/DGA PhD scheme.}
}
\markboth{}%
{Shell \MakeLowercase{\textit{et al.}}: \xr Computed Tomography Reconstruction with Generative Adversarial Networks using Shape Priors}
\maketitle

\begin{abstract}
Compensating scarce measurements by inferring them from computational models is a way to address ill-posed inverse problems. We tackle Limited Angle Tomography by completing the set of acquisitions using a generative model and prior-knowledge about the scanned object.
Using a Generative Adversarial Network as model and Computer-Assisted Design data as shape prior, we demonstrate a quantitative and qualitative advantage of our technique over other state-of-the-art methods. Inferring a substantial number of consecutive missing measurements, we offer an alternative to other image inpainting techniques that fall short of providing a satisfying answer to our research question: can X-Ray exposition be reduced by using generative models to infer lacking measurements?  
\end{abstract}


\IEEEpeerreviewmaketitle

 \section{Introduction}

\subfile{sections/introduction}

\section{Proposed Approach}

\subfile{sections/proposedApproach}

\section{Datasets}

\subfile{sections/Datasets}

\section{Implementation Details}

\subfile{sections/implementationDetails}

\section{Results}

\subfile{sections/results}

\section{Discussion and Conclusion}

\subfile{sections/conclusion}

\section{Bibliography}

\subfile{sections/bibliography}
\section{Tables}

\subfile{sections/tables}

\section{Images}

\subfile{sections/images}

\end{document}

%% file: sections/commands.tex
\newcommand{\xr}{X-Ray }
\newcommand{\xrs}{X-Rays }

%% file: sections/introduction.tex
\IEEEPARstart{X}{-Ray} Computed Tomography (XCT) is a versatile 3D imaging technique that allows the estimation of volumetric \xr attenuation profiles. It produces cross-sectional images of bodies sensitive to this radiation by sampling an object from different viewing angles to reconstruct an image from this sequence of acquisitions. Yet, \xrs are toxic for in-vivo diagnosis and time-consuming in industrial testing. There is a trade-off between sufficient sampling for high-quality images and \xr intake for time and health constraints. Can computational methods exploit prior knowledge about the scanned object to compensate scarce acquisitions by inferring measurements?

We use Generative Adversarial Networks (GAN) to complete the sequence of scarce acquisitions by inferring them from Computer-Assisted Design data. When imaging a slice through an object from the viewing angle $\theta$, with a detector with $R$ pixels, the measurement can be described as
\begin{equation*}
F(\theta,r) = \int_{x} \int_{y}  f(x, y) \delta(x \cos(\theta) + y \sin(\theta) -r )  \ dx \ dy
\end{equation*}
with $r$ the index of pixels on the detector, $(x,y)$ the coordinates of the object and $f$ the density function of the object, that maps a spatial position to a material density. Let $\Theta$ be the set of viewing angles at which the object is sampled. The set of measurements
\begin{equation*}
	S=\{F(\theta,r)\} \textrm{ for all } \theta \in \Theta
\end{equation*} 
is the sinogram of the image $f(x, y)$. It can be represented as an image of size $\vert \Theta \vert \times R$. As such, missing acquisition in the sequence is represented as zero-valued pixels along one dimension of the sinogram. The problem of inferring missing acquisitions from a scarce sinogram is then similar to the one of inferring arbitrarily large regions in images based on image semantics, known as semantic inpainting.

Semantic image inpainting is a constrained image generation problem \cite{pathak2016context}. Missing parts of an image are inpainted using a generative network and solving an optimisation problem. GAN \cite{goodfellow2014generative} and their fully-convolutional version Deep-Convolutional GAN (DCGAN) \cite{radford2015unsupervised} are adapted to this task: they were used for image \cite{yeh2016semantic} and sinogram \cite{yoo2019sinogram} inpainting. The optimisation relies on finding the "closest" encoding in the latent space of the GAN distribution by minimising a penalty function that encompasses contextual and conditional information. This method suffers from several limitations:
\begin{itemize}
	\item Walking the latent space of the distribution can only yield certain improvements in the image generated by the GAN: \cite{jahanian2019steerability} shows that images can only be transformed to some degree (brightness, zoom, rotation). Not only is the transformation corresponding to inpainting is not defined, but it has no certainty to be achievable by "steering" the generated output, especially when guided by a generic loss function and not by a supervised walk.
	\item The optimisation function adds computational time and hyperparameters. In addition to training the generative model, the optimisation process is time-consuming and also requires fine-tuning of the learning rate and number of iterations. 
\end{itemize}
As an alternative to this process, we use CAD data as a prior and train a Unet-GAN \cite{ronneberger2015u}, \cite{isola2017image} to infer missing parts of the sinogram given shape information about the scanned object.

Shape information is often available in both medical and industrial imaging, but is rarely used in XCT reconstruction. For instance, in medical imaging, projects such as \cite{ackerman1998visible} and \cite{xu2007boundary} demonstrate the potential of using numerical shape models of a generic human body to minimise \xr exposure. An alternative is to extract prior information from earlier scans \cite{huang2013iterative, abbas2013super}. For manufactured components, Computer Assisted Design (CAD) drawings are often available, providing strong constraints on object shape. These types of priors provide estimates of object boundary locations, though might not contain information on exact \xr absorption within an object, nor do they contain information about unknown defects and inclusions. 

In this paper, we minimise \xr intake by inpainting the scarce sinogram with a GAN and a shape prior. The main advantage compared to other inpainting methods is the side-stepping of the optimisation process. Unlike the other image inpainting methods, we focus on inpainting the missing part of the sinogram only, reducing the complexity of the task. Our main findings are:
\begin{itemize}
	\item Exploiting the specificity of XCT data enhances the CAD prior. Before feeding the CAD to the GAN, we rescale its values so that they match what has been observed in the scarce sinogram.
	\item Prior information about the shape of the object improves significantly the quality of state-of-the-art (SOTA) sinogram-enhancing techniques. We show that including the CAD prior to other methods that address a similar problem yields a significant improvement of their performance.
	\item Not only does our method enhance the sinogram in terms of Peak Signal-to-Noise Ratio (PSNR) and Structural SIMilarity (SSIM)compared to other SOTA techniques, we also report an improvement of the reconstructed image quality.
\end{itemize}

%% file: sections/proposedApproach.tex
Our method is focused on inpainting a scarce sinogram using information from a shape prior, instead of an optimisation process. As such, it requires an initial training of the generative model, in our case a GAN, followed by the inpainting process. The training procedure is visually detailed in Fig. \ref{fig:training_procedure_explanation}.

\subsubsection{GAN introduction}
A GAN is made of two networks that compete against each other. The generator \textit{G} tries to generate samples that follow the same distribution $p_{data}$ as the training examples by up-sampling a noise vector drawn from a noise distribution $p_{z}$. The discriminator \textit{D}, tries to discriminate between  real training samples and those generated by \textit{G}. GANs have been proposed as an alternative to avoid difficulties commonly found in deep generative models, such as explicit density estimation. GAN optimisation is done by solving the minimax problem: 
\begin{align*}
\label{GAN_Loss}
\min_G \max_D \mathcal{L}(\mathit{G}, \mathit{D}) &= \mathbb{E}_{x \sim p_{data}} \log(D(x)) +\\
& \mathbb{E}_{z \sim p_{z}} \log(1-D(G(z)))
\end{align*}
where $\mathbb{E}$ is the expectation over the training dataset. This architecture uses fully-connected units, which limit the maximum size of images the GAN can generate. To scale to larger images, the Deep Convolutional GAN (DCGAN) architecture \cite{radford2015unsupervised} was proposed. 

\subsection{Deep Convolutional GAN and Unet - The pix2pix Architecture}

GAN are notoriously difficult to train \cite{goodfellow2016nips} and do not handle prior-knowledge as they sample a latent distribution. To address these issues, \cite{isola2017image} proposes the pix2pix architecture that combines the DCGAN improvements and includes shape priors by changing the generator into a U-net architecture.

The main change offered by the DCGAN is the replacement of fully-connected units by convolutional layers. Other modifications include the replacement of the maxout \cite{goodfellow2013maxout} activation by ReLu and Tanh in the generator and LeakyReLu in the discriminator, the inclusion of batch-normalisation \cite{ioffe2015batch} and the replacement of pooling units with learnt sampling units. 

U-nets were initially designed for image segmentation. They are fully convolutional auto-encoders that have residual connections between the down-sampling and up-sampling units. They are convenient architectures that can map a multi-channel input image to a one-channel output image, or vice versa. The pix2pix architecture makes use of this design to add prior-knowledge to the GAN. 

\subsection{Adapted Loss Function}

To encourage the generator to produce images close to their target value, in the sense of the L1-loss, an additional term is added to the training loss as prescribed in \cite{pathak2016context}. We use the L1-Loss  which is a good choice for image quality when Poisson noise is present, as is the case in XCT images \cite{dupe2011inverse}. Unlike other inpainting processes, we train the network to infer only the missing part of the image. We thus use the loss function:
\begin{align}
\min_G \max_D \mathcal{L}(\mathit{G}, \mathit{D}) &= \mathbb{E}_{x \sim p_{true}} \log(D(x)) +  \\
& \mathbb{E}_{\tilde{x} \sim p_{impaired}} \log(1-D(G( \tilde{x}, y ))) + \\
& \lambda L_1 (G( \tilde{x}, y ), x) 
\end{align}
 
where $x$ and $\tilde{x}$ are the  sinograms with all and missing data respectively, $y$ is an image that encodes the shape prior and $\lambda$ a weighting parameter. Given a sinogram and a shape prior, $\mathit{D}$ determines if the sinogram was generated by the GAN or drawn from the true dataset. Given a sinogram with missing data, $\mathit{G}$ generates missing acquisitions that are close to the ground-truth acquisitions in terms of L1-Loss\cite{larsen2016autoencoding}.

\subsection{Architectural Details}
We follow the network's description given in \cite{isola2017image}. Let Ck be a Convolution-BatchNorm-ReLU layer with k filters. Let CDk be a Convolution-BatchNorm-Dropout-ReLU layer with a dropout rate of 50\% \cite{srivastava2014dropout}. All convolutions are 4 by 4 spatial filters applied with a stride of 2. Convolutions in the encoder down-sample by a factor of 2, whereas in the decoder they up-sample by a factor of 2. After the last layer in the decoder, a convolution is applied, followed by a Tanh function and an element-wise multiplication with a mask to infer missing acquisitions only, as explained in \ref{encoding_the_shape_prior}. As an exception to the above notation, BatchNorm is not applied to the first C64 layer in the encoder. All ReLUs \cite{nair2010rectified} in the encoder are leaky \cite{maas2013rectifier}, with slope 0.2, while ReLUs in the decoder are not leaky. This results in the following generator:

\textbf{encoder:}

C64-C128-C256-C512-C512-C512-C512-C512

\textbf{decoder:}

CD512-CD1024-CD1024-C1024-C1024-C512-C256-C128

The discriminator works on image patches of size 16x16. For the discriminator, after its last layer and after extracting the patch, a convolution is applied to map to a one-channel output, followed by a Sigmoid function. This results in the following discriminator:

\textbf{encoder:}

C64-C128-C256-C512

\subsection{The Inpainting Process}
Once the network is trained and its parameters frozen, the inpainting process is straightforward. It requires a scarce sinogram, a sufficiently sampled CAD sinogram and a mask indicating the acquisitions to infer. Given this input data, $\mathit{G}$ infers the missing acquisitions in a one-channel image. This image is then added pixel-wise to the scarce sinogram to produce the inpainted sinogram. A visual explanation of the method is given in Fig. \ref{fig:inpainting_method_explanation}.

\subsection{Other Methods for Comparison}
An immediate solution to inferring missing acquisitions is to use linear interpolation. Using the shape prior, replacing the missing acquisitions by the ones expected by the CAD is another solution. An improvement of the acquisition prior is brought by scaling it with attenuation values matching the ones seen in the measured sinogram: this scaling is detailed in \ref{encoding_the_shape_prior}. We here implement these three solutions for comparison.

The pix2pix architecture used here has never been used for XCT data but other similar methods exist. Indeed, \cite{lee2018deep} replaces missing acquisitions by linearly interpolated ones and then uses a Unet \cite{ronneberger2015u} to enhance their quality. We implement this architecture without the shape prior as in the original paper and also produce a version that includes our shape prior to make it more comparable to our new approach. A GAN architecture is also used for sinogram synthesis \cite{yoo2019sinogram}, with an optimisation procedure to perform the inpainting. This model does not use shape priors. We implemented this architecture and trained it with as much care as we trained the other, but could not obtain a satisfactory result.

%% file: Sinogram Enhancement with Gans/sections/datasets.tex
We tackle the sinogram inpainting with shape information problem using a data-driven method, that relies on large datasets. However, such datasets of XCT acquisitions with associated shape priors do not exist. We therefore adapted the SophiaBeads \cite{coban2015sophiabeads_dataset} dataset to report the performance of our process on real data. 

\subsection{Preprocessing the SophiaBeads Dataset}
Initially designed to provide data for researchers in XCT reconstruction to benchmark their algorithms, this dataset is a collection of six cone beam XCT acquisitions of a plastic tube filled with soda-lime glass beads. The difference between each dataset is the number of projections per volume. As the beads are approximately circular, each slice through the volume contains only circular objects.

We reconstruct two high-projections datasets using the approach in \cite{coban2015sophiabeads_codes} and \cite{biguri2016tigre}, generating 256x256x200 volumes. For experimental convenience and speed, training data is generated from individual 2D slices, using a parallel beam geometry. Circle finding methods are used to identify object boundaries, which were used as shape priors.  

Thus, our datasets consist of two volumes of 200 slices each, together with the associated CAD data. Each volume is split into 180 training slices and 20 test slices. To test the robustness of our method, the outline of the plastic container was not captured in the CAD data and several small circles are also not included. This is done to ensure that the images had features that are not represented in the CAD data. A slice of the volume reconstructed with 512 projections and the identified beads, from which the object boundaries can be inferred are shown in Fig. \ref{fig:sophiabeads_obj_slice} and Fig. \ref{fig:sophiabeads_cad_slice}, respectively.

We generate full sinograms from each slice with \cite{biguri2016tigre}, simulating a detector with 256 pixels and measuring from 256 equally spaced angles in a 180 degree arc. The sinograms corresponding to Fig. \ref{fig:sophiabeads_obj_slice} and Fig. \ref{fig:sophiabeads_cad_slice} are shown in Fig. \ref{fig:sophiabeads_obj_sin} and Fig. \ref{fig:sophiabeads_cad_sin}, respectively. Sinograms with missing projections are generated by setting between 5 to 95\% of consecutive angular measurements to 0. It is worth emphasizing that we did not remove scan directions uniformly at random, but choose the more challenging setting where larger numbers of consecutive angles are missing. An example of a sinogram with half of its data missing is shown in Fig. \ref{fig:sophia_sin_example}.

\subsection{Synthetic Training Data}
\label{synthetic_training_data}

Having a synthetic dataset allows experimenting with the data model. In real applications, the shape prior will not always exactly match the object shape. There could be boundary deviations, or objects could have unknown inclusions and internal defects. Therefore, we also generate separate data with additional air pockets in the objects that were not captured by the shape prior, as can be seen in Fig. \ref{fig:holes_examples}. 

In order to simulate such defects, we create a synthetic dataset to mirror the SophiaBeads data, as shown in Fig. \ref{fig:syn_obj_slice} and Fig. \ref{fig:syn_cad_slice}. Each slice is a 256 by 256 image of several circles generated with the \textit{circle} function from the scikit-image.draw Python module with different radii and positions and uniform densities with added Gaussian noise. Object attenuation values are set to 25 while the image background is assumed to have an X-Ray attenuation much lower than the object and is set to 1. The shape prior associated to a slice is generated using the same function with the same radii and position for each circle but with different densities and no Gaussian noise. Whilst we chose this model to replicate the SophiaBeads dataset, similar data has previously been used in \cite{jin2017deep}, \cite{adler2017solving} and \cite{kelly2017deep}.

%% file: Sinogram Enhancement with Gans/sections/ImplementationDetails.tex
Our approach relies on the use of a shape prior to train a generative model. The encoding of this prior and the training of the model are then crucial in the process.

\subsection{Encoding the Shape Prior}
\label{encoding_the_shape_prior}

There are several ways to encode the object's shape information for use by the GAN. For instance, one can generate images where the boundary pixels are set to a non-zero value. Also, the object boundary pixels and the object interior pixels can be set to a single value, using the same value for all objects. Otherwise, we set the object boundary pixels and the object interior pixels to a random value, using the same value for one object, but different values for different objects. We chose the latter as it allows a greater range of different densities to be encoded.

To enhance the prior, we add a pre-processing step that makes use of a property of sinograms. For an acquisition angle, when summing the pixel values read on the detector, one gets the sum of all the attenuation values on the object, i.e. 
\begin{equation}
\int_r R(\theta,r) = \int_y \int_x f(x,y) \ dx \  dy = \mathrm{const}.
\end{equation} 
This value is constant for all $\theta$ if all objects remain entirely within the field of view. Here, we use this estimate of the overall attenuation to scale the sinograms derived from the shape prior encoding. We also define a binary sinogram mask that indicates sinogram pixels where we know that information is missing, i.e. those pixels for which $\theta$ is in the set of missing pixels but where the shape prior predicts a non-zero sinogram value, as can be seen in Fig. \ref{fig:sophia_sin_mask}. We concatenate all three modalities into a three channel image as the input to the GAN generator.
 
\subsection{GAN Training}

The networks are trained on the SophiaBeads training dataset reconstructed with 1024 projections, using a hundred epochs. They are then evaluated on the dataset reconstructed from 512 projections. When using prior information, we concatenate it to the sinogram before training the networks with it. The concatenated input data is re-scaled to have values between 0 and 1. We used a batch size of 8 and the Adam optimizer \cite{kingma2014Adam}. The learning rate is set at 0.0002 and $\lambda$ is set to 100. We used label-smoothing and noise addition in the discriminator, as prescribed in \cite{goodfellow2016nips}. We have implemented the top-k \cite{sinha2020top} training procedure which discards 6/8th of the generated samples when updating the \textit{G} loss. We trained comparison networks with the same procedure but, as no information was given in \cite{yoo2019sinogram} and \cite{yeh2016semantic} on the hyper-parameters of the inpainting step used in their methods, our implementation of these did not produce satisfactory results. All of the code used in this paper are available at \cite{valat2021codes}, the implementation is made using PyTorch.

Training is performed on the SophiaBeads dataset with the priors and on a synthetic dataset that models internal defects separately. It means that two copies of the same architecture were trained separately on each dataset, but with the same hyper-parameters.

%% file: sections/results.tex
\label{sec:results}
Even if the choice of evaluation metrics for synthetic images remains an open problem \cite{salimans2016improved}, the evolution of well-known loss functions is a good estimate of the effect of the network on the sinogram. Thus, we chose to report the PSNR and the SSIM between predictions and ground-truth. We compare our method to state of the art methods from the literature. In particular, we compare it to the sinogram inpainting method of \cite{yoo2019sinogram} and the U-net approach in \cite{lee2018deep}, where we also implemented a version that used additional CAD constraints. These constraints were enforced here by concatenating the scarce sinogram, the CAD sinogram and the mask before training the U-net with it.

\subsection{Estimation Performance on the SophiaBeads Dataset}
We have two volumes of 200 slices each. Each volume is split into 90 percent training set and 10 percent testing set. Training is performed on the training set of the volume reconstructed with 1024 projections and testing is done on the testing set of the volume reconstructed with 512 projections.

\subsubsection{Impact on the sinogram}
\label{impact_on_the_sinogram_sophiabeads}

As can be seen in Table\ref{table:results_sophiaBeads_sinograms}, the pix2pix architecture combined with the CAD prior outperforms all other compared approaches. It is interesting to see that using the CAD prior significantly enhances the performance of the U-net. It is also worth mentioning that the scaling operation slightly improves the quality of the CAD replacement. The inpainting process has limitations that we  will detail in \ref{sec:conclusion}.

\subsubsection{Impact on the reconstructions}
The reconstructions are made using the Simultaneous Iterative Reconstruction Technique algorithm and implemented with TIGRE for Python \cite{biguri2016tigre}.

As can be seen in Table\ref{table:results_sophiaBeads_sinograms}, not only does the pix2pix architecture combined with the CAD prior outperform all other approaches considered in the image space, but it is the only one that yields an improvement compared to the reconstruction made from the scarce sinogram. It is also important to underline that methods performing well in the sinogram space do not necessarily yield an improvement in the image space. This might be due to the reconstruction algorithm used. Fig. \ref{fig:reconstructions} shows the reconstructions associated with each method for 128 missing acquisitions.

\subsection{Effect of Inconsistencies Between the Two Modalities on the Synthetic Dataset}
\label{inconsistencies}
The advantage of our method here is less strong and a qualitative inspection shows that the network fails at inferring the hole pattern. It shows a limited ability to recover defects. In Fig. \ref{fig:images_with_holes}, we show four examples of the difference between the image reconstructed with our method and the target image, for 60 percent missing acquisitions. This indicates that the GAN fails to learn the fine structure of the object's attenuation.

One reason for this might be that we are estimating a very large number of missing projections, which is an extremely challenging task. Also, the nature of holes makes the choice of loss function delicate, as they are "dents" in an acquisition and the GAN produces a high-frequency signal, which is very similar to this pattern. In a future study, we plan to modify the architecture to generate one acquisition at a time, hoping to overcome this limitation. One must also note that the image reconstruction process does not weigh real versus GAN-inferred acquisitions differently, which might improve performance as it would take uncertainty in estimates into account.

%% file: sections/conclusion.tex
\label{sec:conclusion}
One must note that the problem tackled by the Unet-only approach is not exactly the same as ours, as the limited-angle tomography is addressed by removing acquisition at regular angular intervals and training on patches of the sinogram. This method is not designed for inferring acquisitions when a substantial number of them are missing. Indeed, if the patch size is 50 and the number of consecutive missing acquisitions is more than 50, which is only 7\% missing acquisitions in the experimental set-up of the paper, the method cannot work as no true information would be given to the network. We also believe that it was not designed for a use case where all the sinogram was processed at once, even if \cite{lee2018deep} mentions that larger patch size had no issues to be dealt with. Our comment on the GAN-only approach is that the sinograms that are of interest are much smaller in size than the ones we are interested in: 128*128 and 180*180 instead of 256*256 in our case. Their method is directly inspired by \cite{yeh2016semantic}, where results are achieved on 64*64 images and where learning is based on contextual as well as discriminator losses. We think that the GAN inpainting without CAD prior fails at the task because we could not train it satisfyingly using the same procedure as we followed for our architecture. 

In this paper we have shown how to use a GAN to exploit shape prior information to address the image inpainting with edge information problem in XCT. Our experiments demonstrate the significant advantage that our method offers over SOTA methods that do not include shape prior information, for inferring a large number of consecutive missing acquisitions. We also demonstrated that the CAD prior facilitates the GAN training. The main limitation of our approach is the failure at extrapolating faults unexpected by the shape priors. As discussed in \ref{inconsistencies}, we believe that it is due to the fact that all missing acquisitions are generated at once. Also, the image reconstruction process is not studied here but we think that weighting the ground-truth and generated acquisitions could improve the final image quality. We conclude by reflecting on the method used for sinogram enhancement. The term inpainting refers to images and we believe that there is a misuse of this technique for sinograms. Indeed, a sinogram is a sequence of acquisitions that is concatenated into an image for visualisation purposes but it is not an image as the word  is intended for a photograph. We then believe that even if it is a valid way to address the problem of inferring missing acquisitions, it is under-performing. Indeed, no information about the sampling geometry is used, nor about the position of the acquisition to infer. This method also constrains the sinograms as they have to be relatively small, two-dimensional, and have the same number of acquisitions as there are pixel detectors. In a future study, we will focus on inferring one acquisition at a time, accounting for the angular proximity to other acquisitions and specificity of the interpolation of acquisitions in XCT.

%% file: sections/tables.tex
\begin{table}[]
\centering
\label{table:results_sophiaBeads_sinograms}
\setlength{\tabcolsep}{0.75\tabcolsep}
\begin{tabular}{|l|l|l|l|l|l|l|}
\hline
Missing angles & \multicolumn{2}{c|}{30 \textdegree} & \multicolumn{2}{c|}{60\textdegree} & \multicolumn{2}{c|}{90\textdegree}\\
\hline
Method & PSNR & SSIM & PSNR & SSIM & PSNR & SSIM\\ \hline
non-scaled CAD & 22.47 & 0.88 & 19.24 & 0.78 & 17.06 & 0.68\\ \hline
scaled CAD & 23.13 & 0.88 & 21.31 & 0.77 & 19.66 & 0.66\\ \hline
linear interpolation & 23.55 & 0.87 & 22.63 & 0.82 & 20.01 & 0.75\\ \hline
Unet & 9.93 & 0.43 & 8.35 & 0.24 & 7.69 & 0.15\\ \hline
Unet + CAD  & 16.78 & 0.83 & 12.90 & 0.72 & 10.95 & 0.61\\ \hline
GAN inpainting & 6.95 & 0.80 & 3.82 & 0.64 & 1.95 & 0.47\\ \hline
pix2pix inference & \textbf{29.10} & \textbf{0.92} & \textbf{27.76} & \textbf{0.88} & \textbf{27.08} & \textbf{0.83}\\ \hline
\end{tabular}
\caption{Quantitative performance improvement on the sinogram comparison between the implemented methods.}
\end{table}

\begin{table}[]
\centering
\label{table:results_sophiaBeads_reconstruction}
\setlength{\tabcolsep}{0.75\tabcolsep}
\begin{tabular}{|l|l|l|l|l|l|l|}
\hline
Missing angles & \multicolumn{2}{c|}{30 \textdegree} & \multicolumn{2}{c|}{60\textdegree} & \multicolumn{2}{c|}{90\textdegree}\\
\hline
Method & PSNR & SSIM & PSNR & SSIM & PSNR & SSIM\\
\hline
no interpolation & 69.15 & 1.00 & 66.47 & 1.00 & 63.77 & 1.00\\\hline
scaled CAD & 61.33 & 0.99 & 58.28 & 0.99 & 56.28 & 0.99\\ \hline
linear interpolation & 61.23 & 1.00 & 61.76 & 1.00 & 59.37 & 1.00\\\hline
Unet & 52.67 & 0.99 & 51.42 & 0.99 & 51.43 & 0.99\\\hline
Unet + CAD & 60.35 & 1.00 & 56.50 & 1.00 & 55.12 & 0.99\\\hline
GAN inpainting & 53.41 & 1.00 & 51.40 & 0.99 & 50.90 & 0.99\\\hline
pix2pix inference & \textbf{69.59} & 1.00 & \textbf{67.89} & 1.00 & \textbf{66.46} & 1.00\\\hline
\end{tabular}
\caption{Quantitative performance improvement on the reconstruction comparison between the implemented methods.}
\end{table}

%% file: sections/images.tex
\begin{figure*}
{\includegraphics[width=0.5\textwidth]{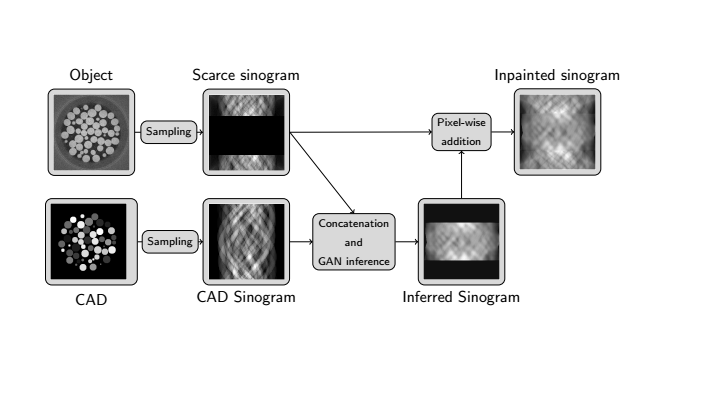}}
\caption{Visual explanation of the inpainting method.}
\label{fig:inpainting_method_explanation}
\end{figure*}

\begin{figure*}
{\includegraphics[width=0.5\textwidth]{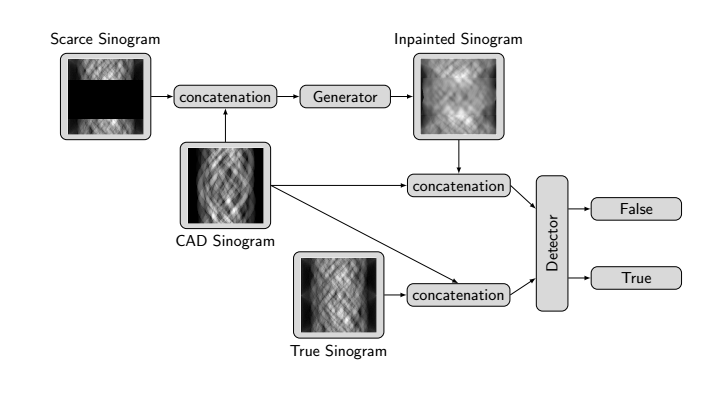}}
\caption{Visual explanation of the training of the GAN.}
\label{fig:training_procedure_explanation}
\end{figure*}

\begin{figure*}
\centering
\subfloat[Slice of the SophiaBeads volume.]{\includegraphics[width=0.2\textwidth]{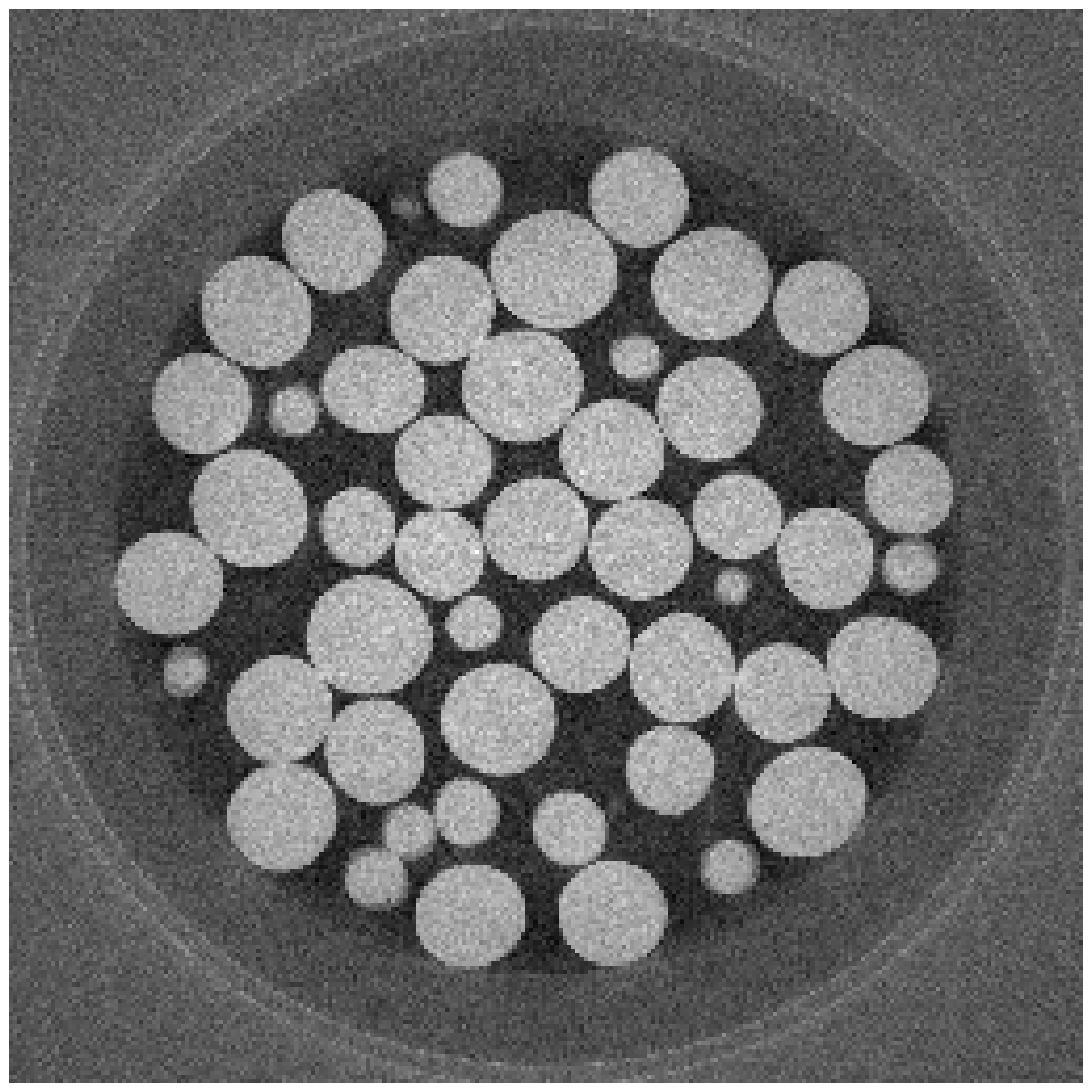}
\label{fig:sophiabeads_obj_slice}}
\hfil
\subfloat[Shape prior associated to Fig. \ref{fig:sophiabeads_obj_slice}.]{\includegraphics[width=0.2\textwidth]{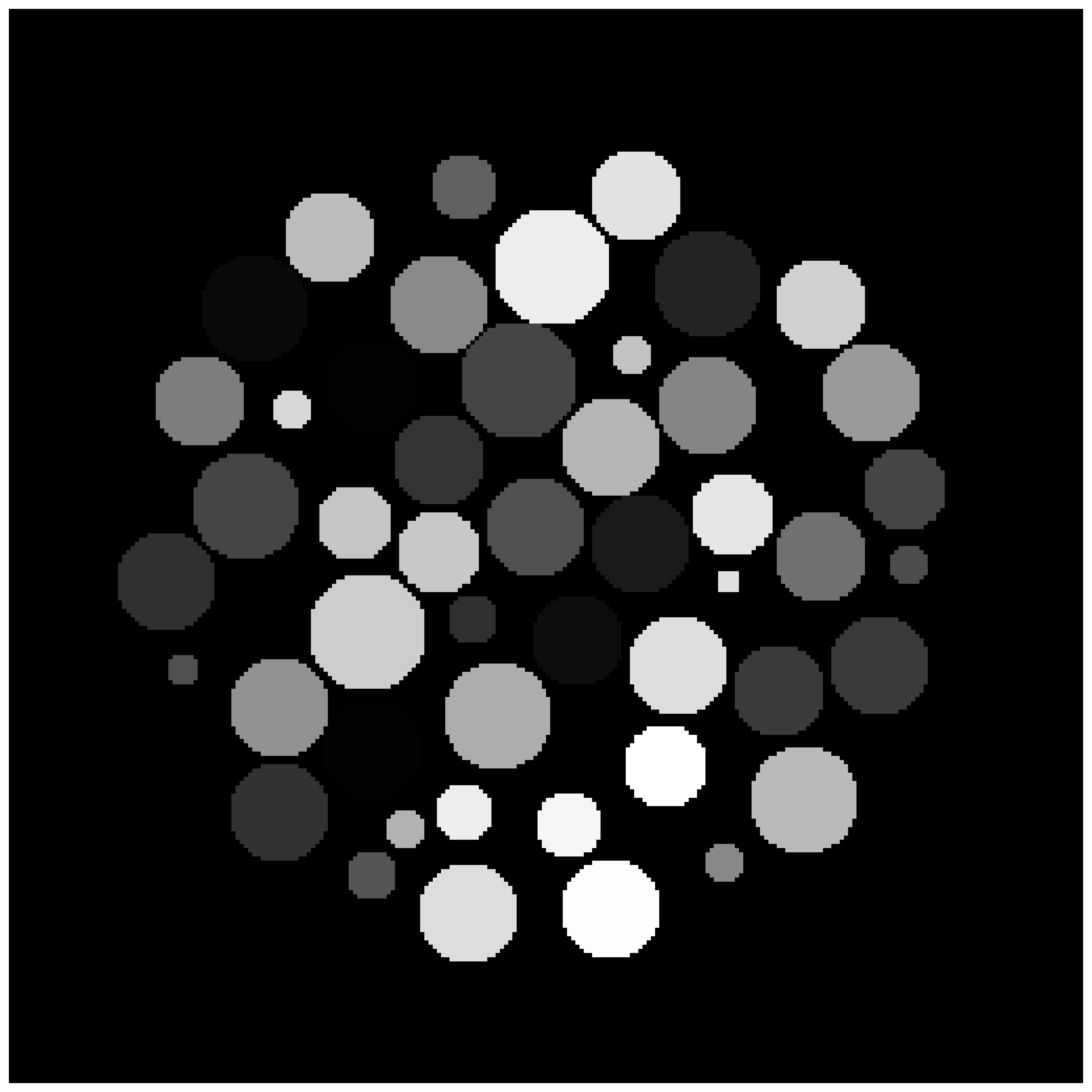}
\label{fig:sophiabeads_cad_slice}}
\hfil
\subfloat[Sinogram corresponding to Fig. \ref{fig:sophiabeads_obj_slice}.]{\includegraphics[width=0.2\textwidth]{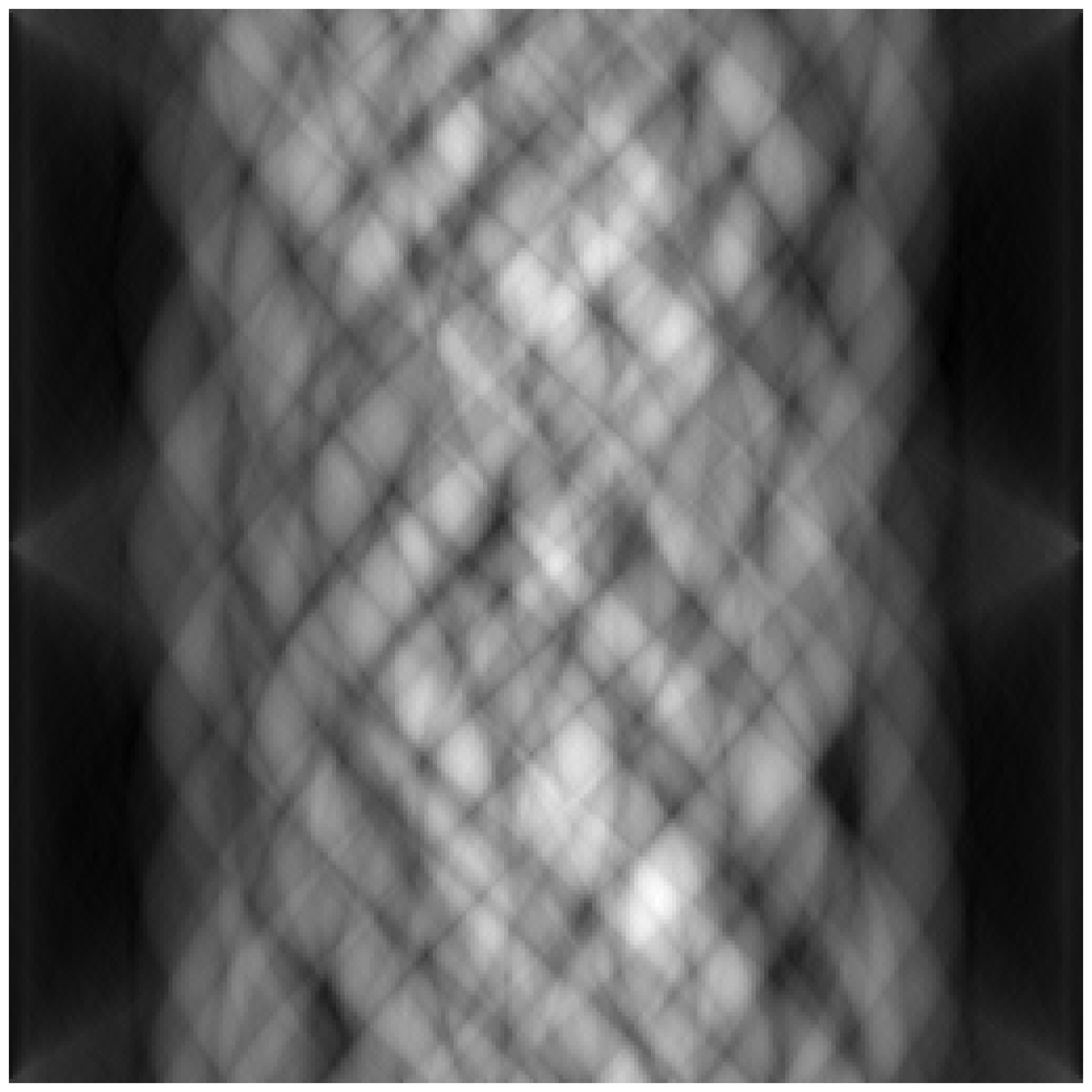}
\label{fig:sophiabeads_obj_sin}}
\hfil
\subfloat[Sinogram corresponding to Fig. \ref{fig:sophiabeads_cad_slice}.]{\includegraphics[width=0.2\textwidth]{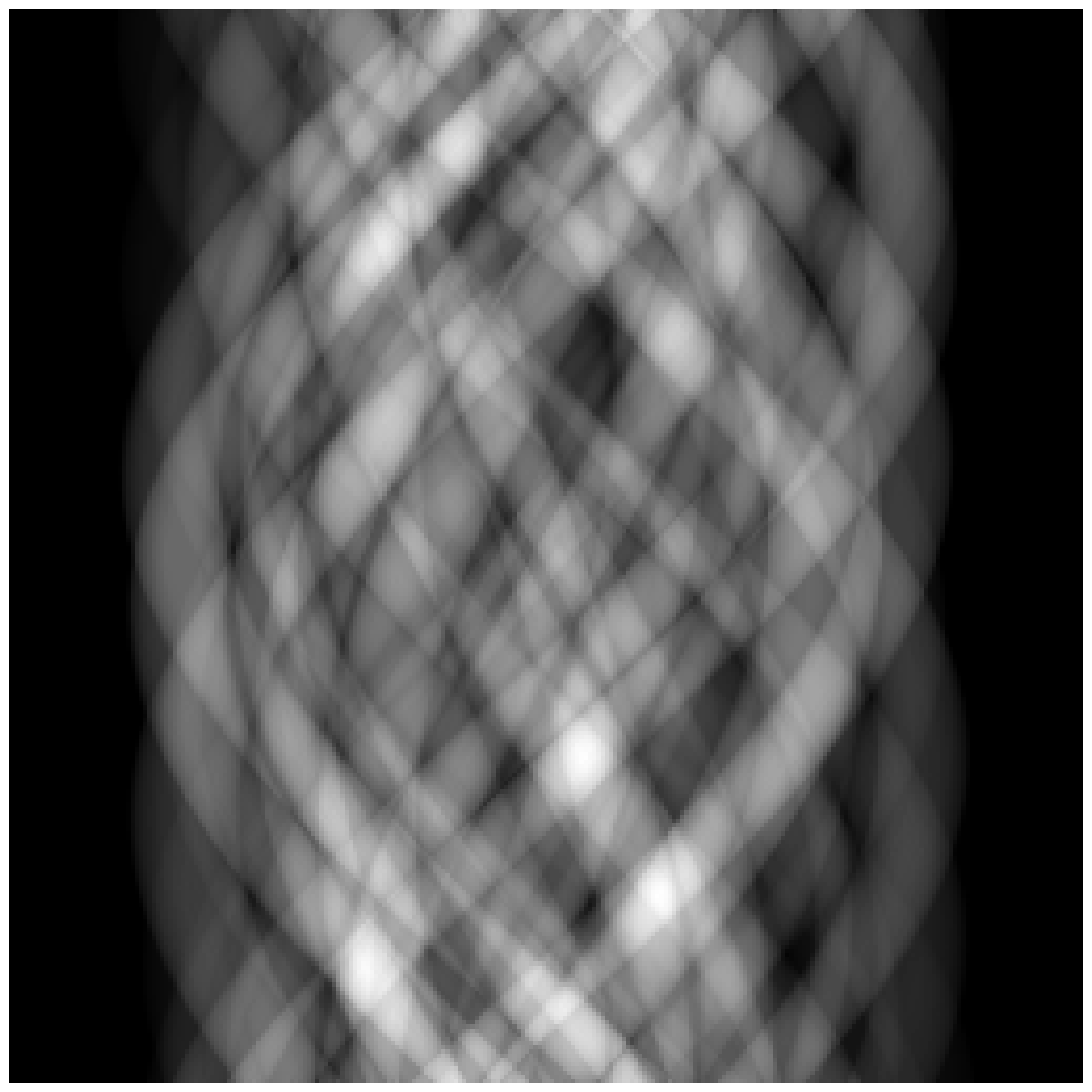}
\label{fig:sophiabeads_cad_sin}}
\caption{Sample from the volume of the SophiaBeads dataset. Fig \ref{fig:sophiabeads_obj_slice} shows the reconstructed acquisition and Fig \ref{fig:sophiabeads_cad_slice} shows a representation of shape priors showing each object with a different randomly assigned gray value, in which no plastic container is visible. The sinograms shown in Fig. \ref{fig:sophiabeads_obj_sin} and Fig. \ref{fig:sophiabeads_cad_sin} are obtained using the forward radon transform with a parallel beam geometry from \ref{fig:sophiabeads_obj_slice} and \ref{fig:sophiabeads_cad_slice}, respectively.}
\label{fig:Sample SophiaBeads}
\end{figure*}

\begin{figure*}
\subfloat[Sinogram with missing acquisitions drawn from the SophiaBeads dataset.]{\includegraphics[width=0.2\textwidth]{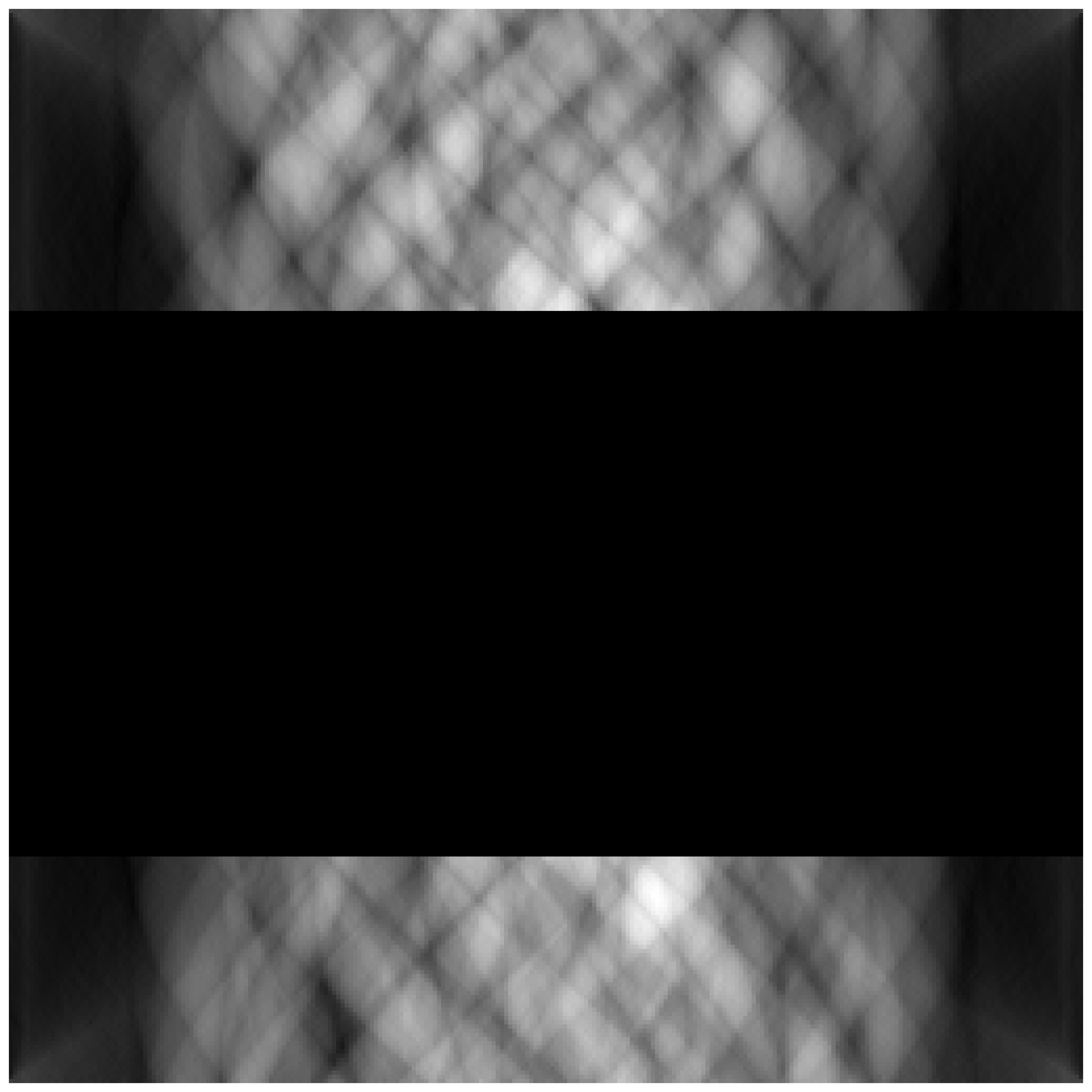}
\label{fig:sophia_sin_example}}
\hspace{5mm}%
\subfloat[Mask associated with Fig. \ref{fig:sophia_sin_example}.]{\includegraphics[width=0.2\textwidth]{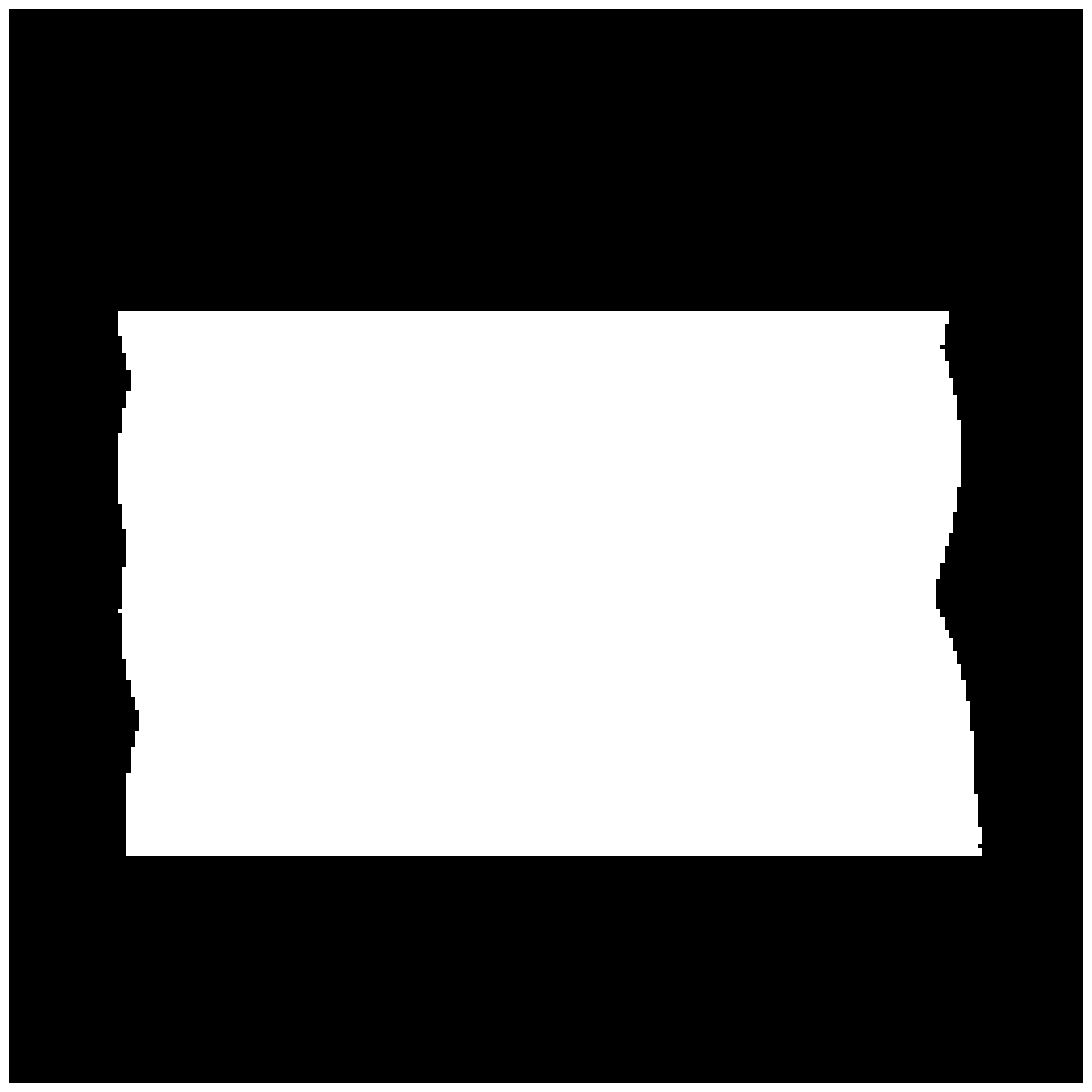}
\label{fig:sophia_sin_mask}}
\caption{Fig. \ref{fig:sophia_sin_example} shows a sinogram with missing acquisitions and Fig. \ref{fig:sophia_sin_mask} its associated mask.}
\label{fig:sinogram_example}
\end{figure*}
    
\begin{figure*}
\centering
\subfloat[Sample from the synthetic dataset.]{\includegraphics[width=0.2\textwidth]{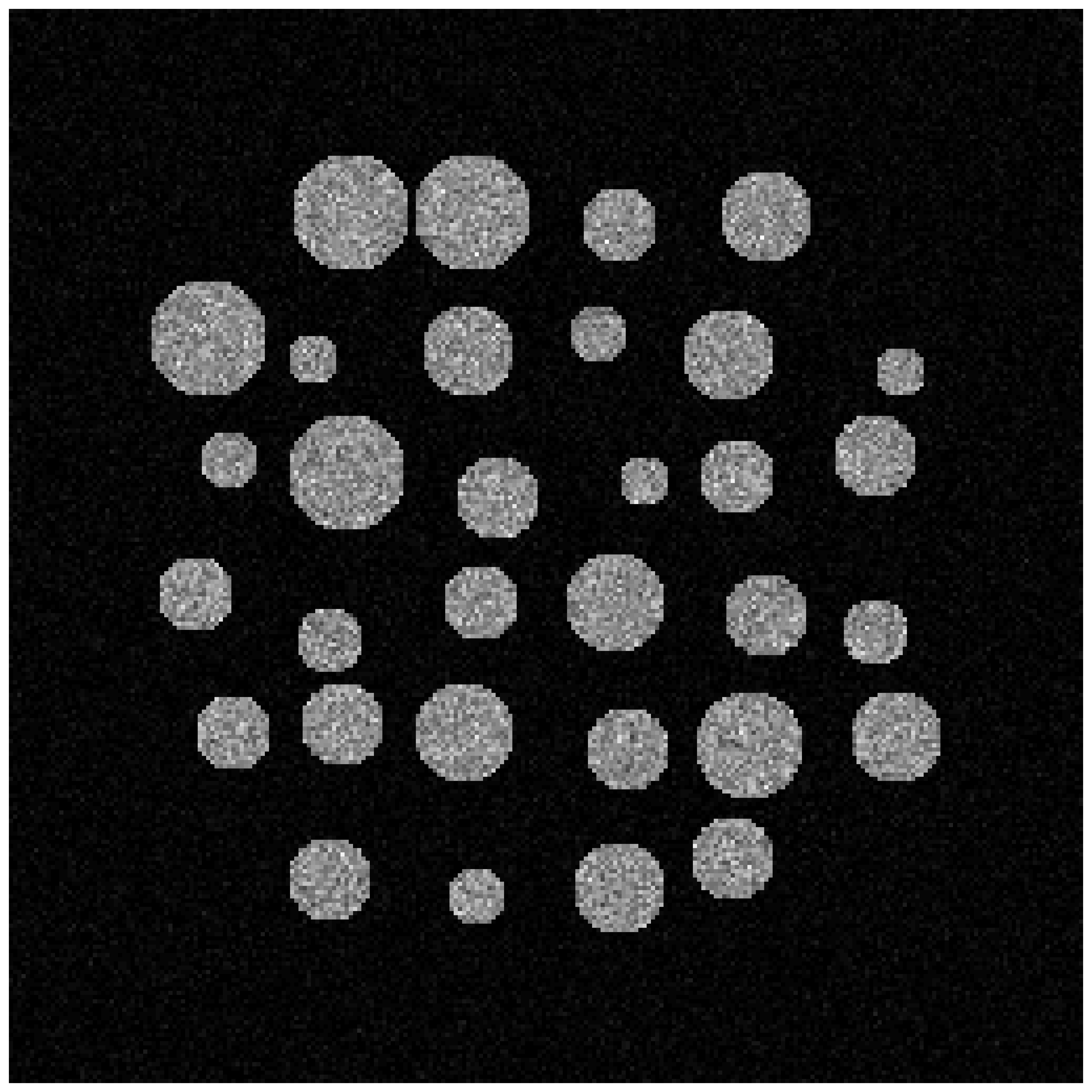}
\label{fig:syn_obj_slice}}
\hfil
\subfloat[Encoded shape prior associated to Fig. \ref{fig:syn_obj_slice}.]{\includegraphics[width=0.2\textwidth]{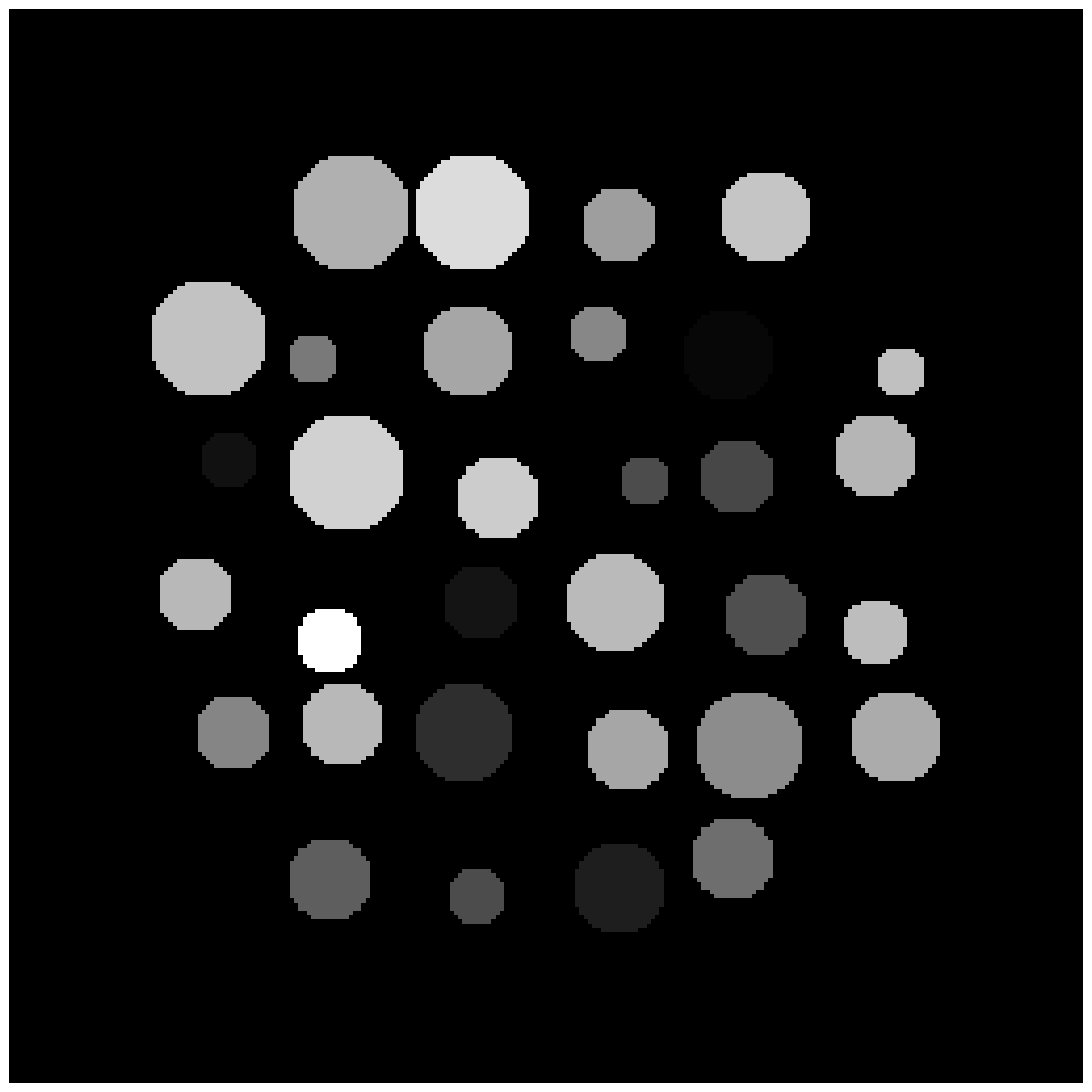}
\label{fig:syn_cad_slice}}
\hfil
\subfloat[Sinogram corresponding to Fig. \ref{fig:syn_obj_slice}.]{\includegraphics[width=0.2\textwidth]{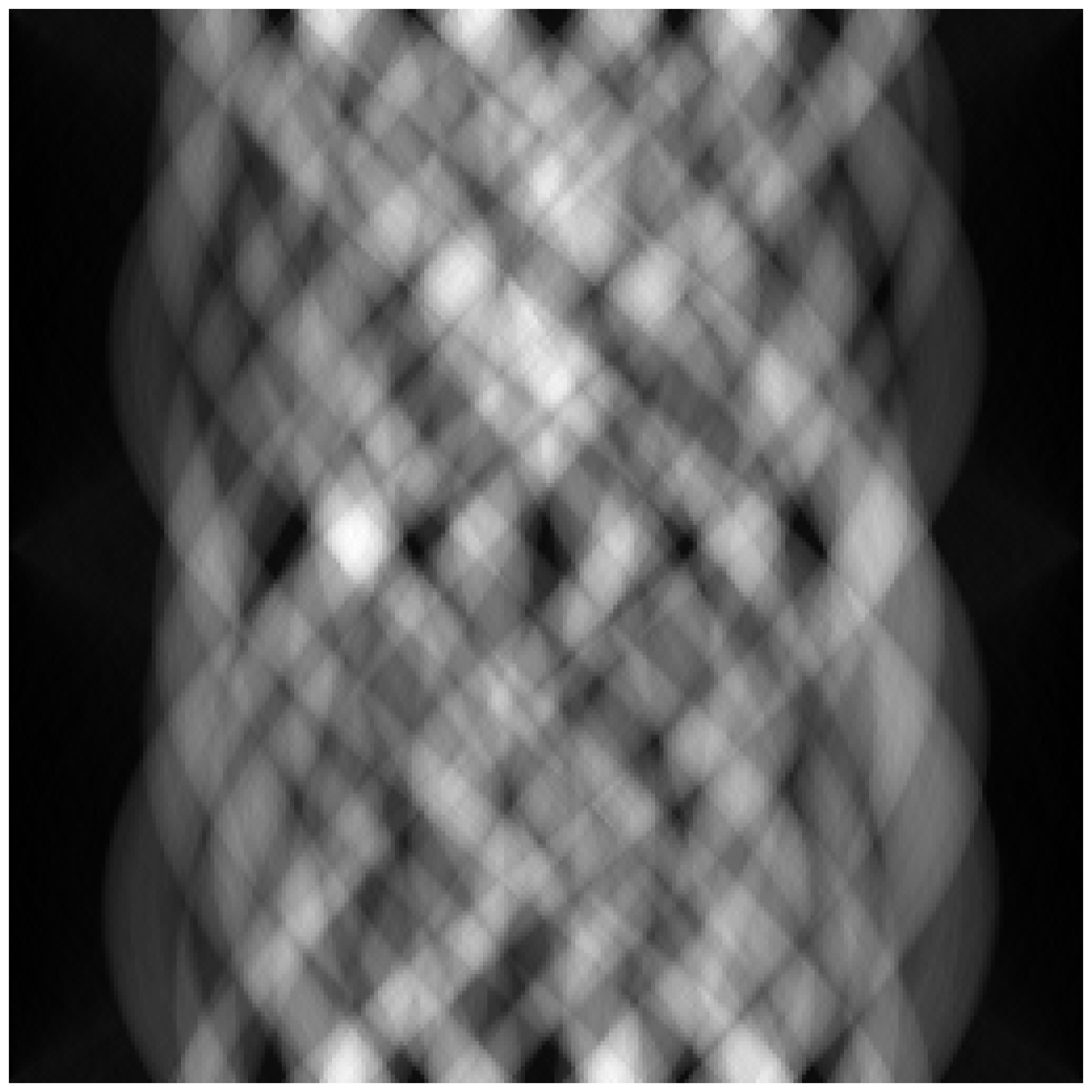}
\label{fig:syn_obj_sin}}
\hfil
\subfloat[Sinogram corresponding to Fig. \ref{fig:syn_cad_slice}.]{\includegraphics[width=0.2\textwidth]{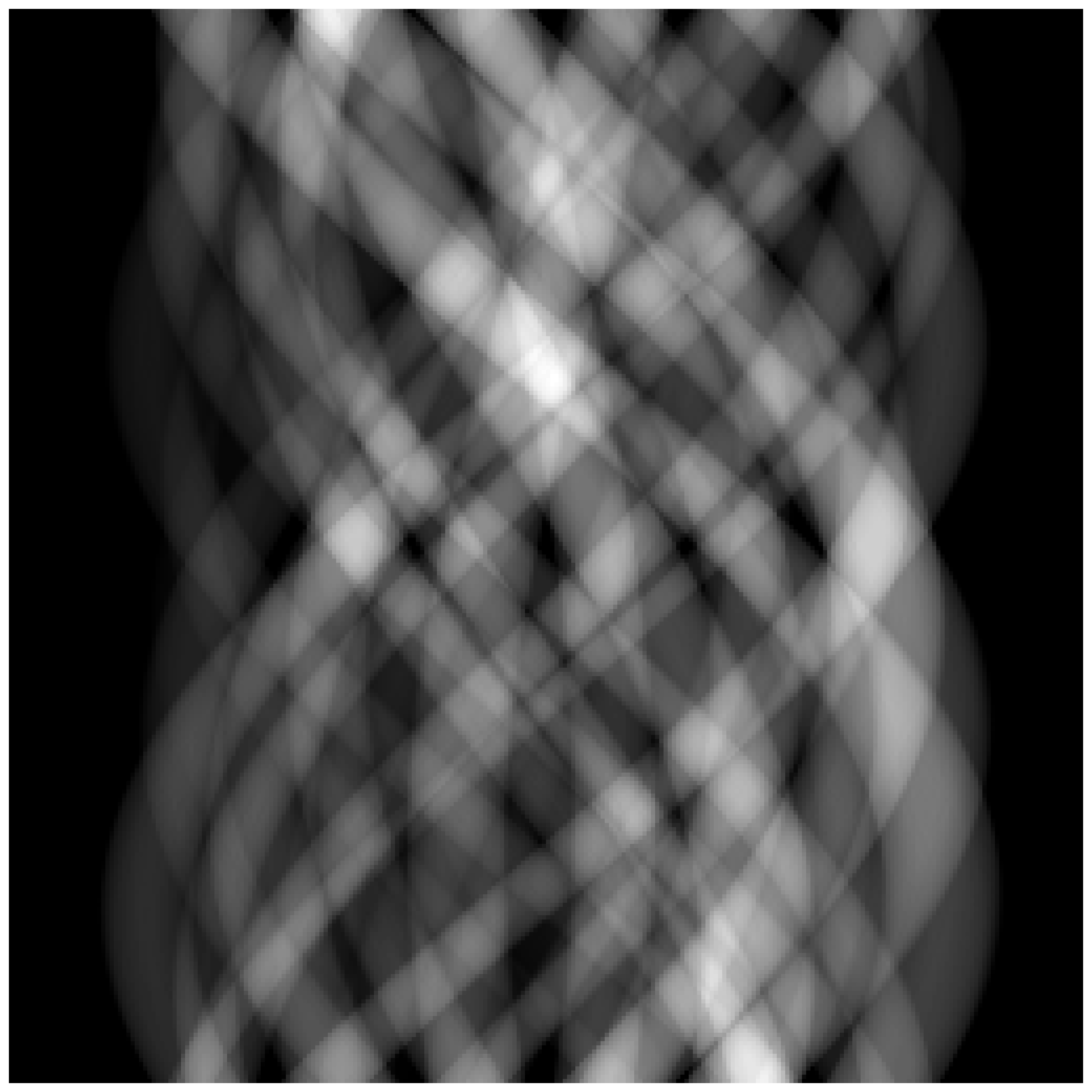}
\label{fig:syn_cad_sin}}
\caption{Sample from the volume of the synthetic dataset. (\ref{fig:syn_obj_slice}) shows the ground-truth reconstruction and (\ref{fig:syn_cad_slice}) shows a representation of the shape priors showing each object with a different randomly assigned gray value, in which no plastic container is visible. In an attempt to reproduce noise in the image, a Gaussian noise is added to the material of uniform densities. The sinograms shown in (\ref{fig:syn_obj_sin}) and (\ref{fig:syn_cad_sin}) are obtained using the forward radon transform with a parallel beam geometry.}
        \label{fig:Sample synthetic}
\end{figure*}          

\begin{figure*}
\subfloat[Example of a sample of the synthetic dataset when internal defects are present.]{\includegraphics[width=0.2\textwidth]{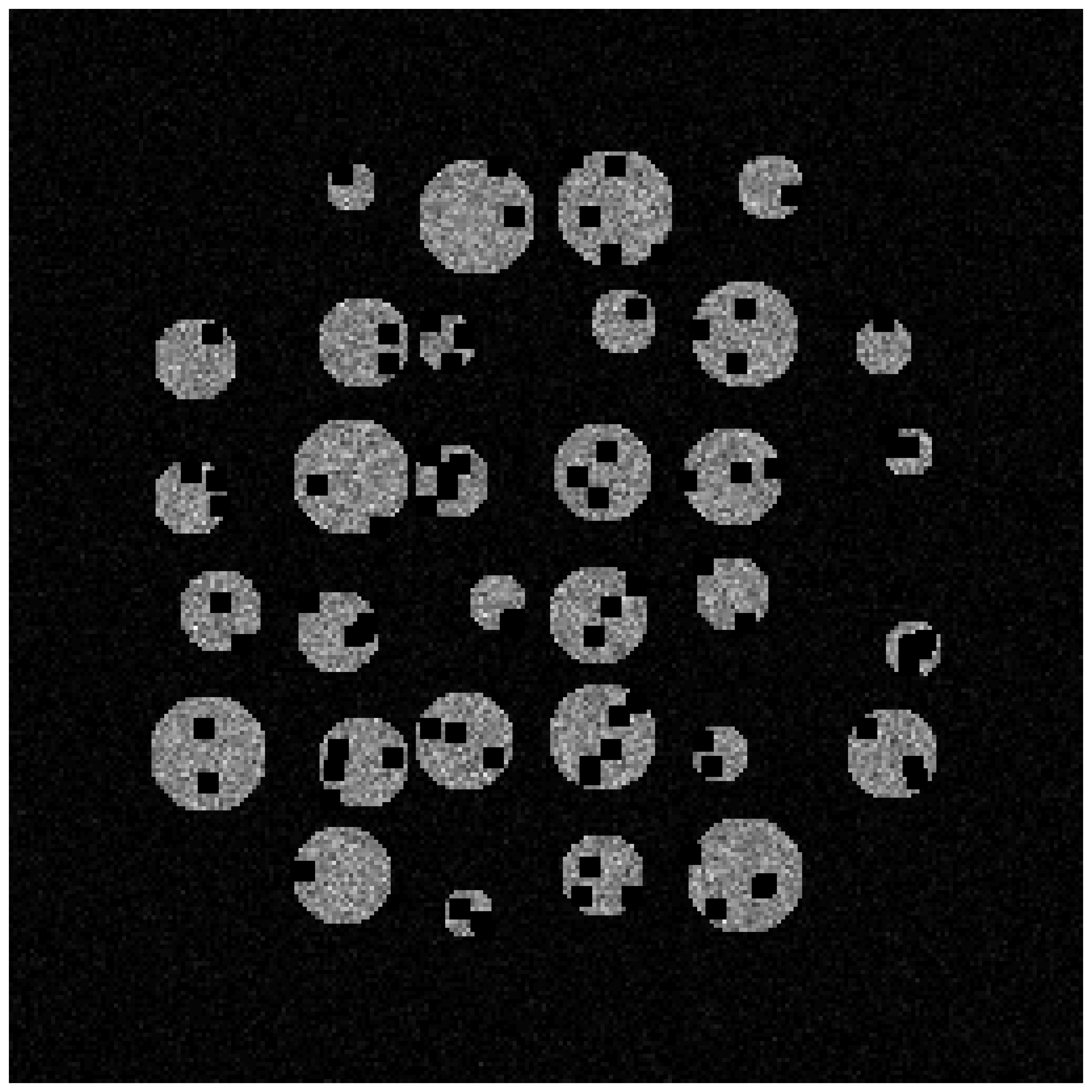}
\label{fig:holes_examples_}}
\caption{Objects with internal defects not encoded in the shape prior.}
\label{fig:holes_examples}
\end{figure*}

\begin{figure*}
\centering
\subfloat[Target reconstruction with 256 acquisitions]{\includegraphics[width=0.2\textwidth]{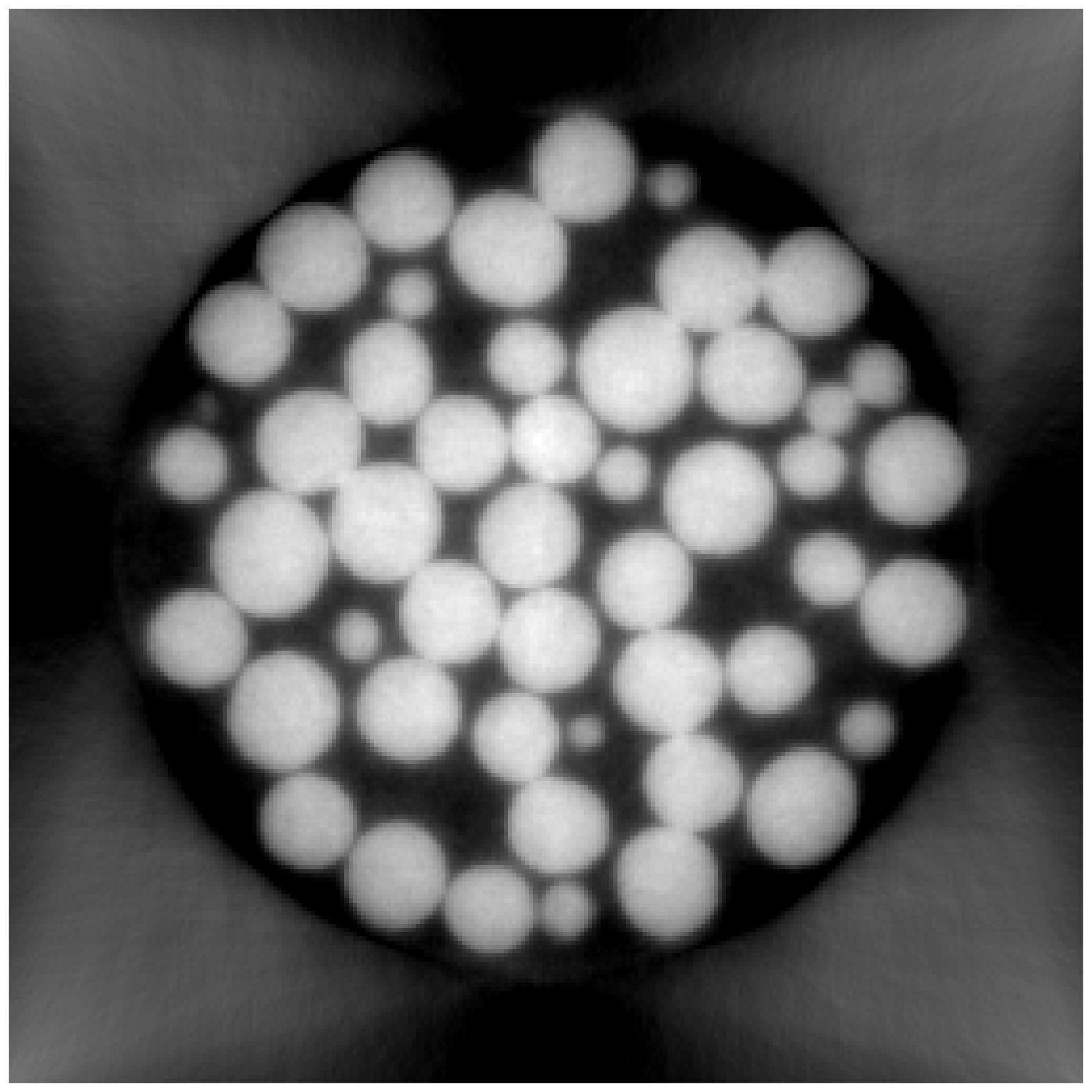}
\label{fig:reconstructions128_target}}
\hfill
\subfloat[Reconstruction without any interpolation method , i.e from 128 acquisitions]{\includegraphics[width=0.2\textwidth]{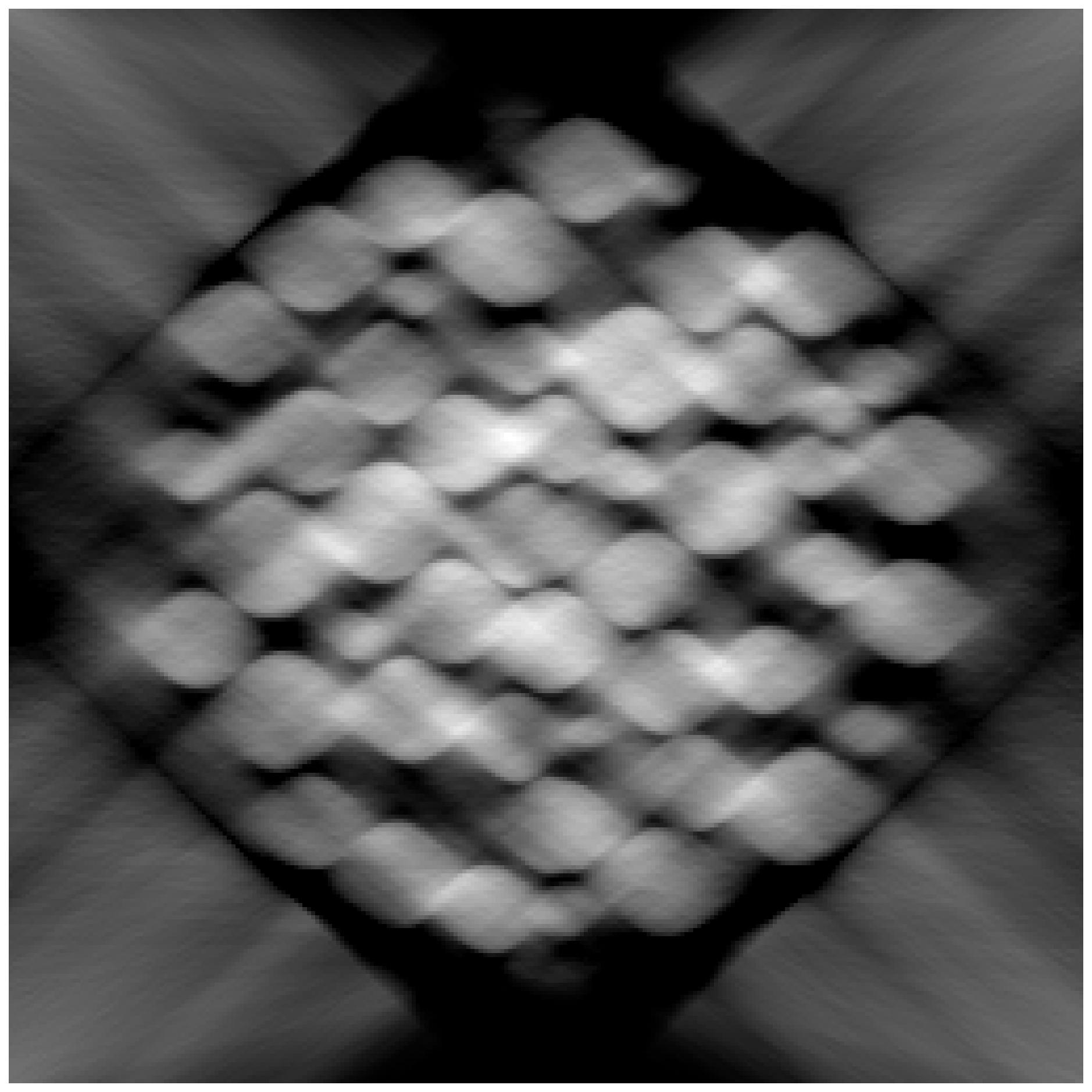}
\label{fig:reconstructions128_noInterpolation}}
\hfill
\subfloat[Reconstruction with linear interpolation]{\includegraphics[width=0.2\textwidth]{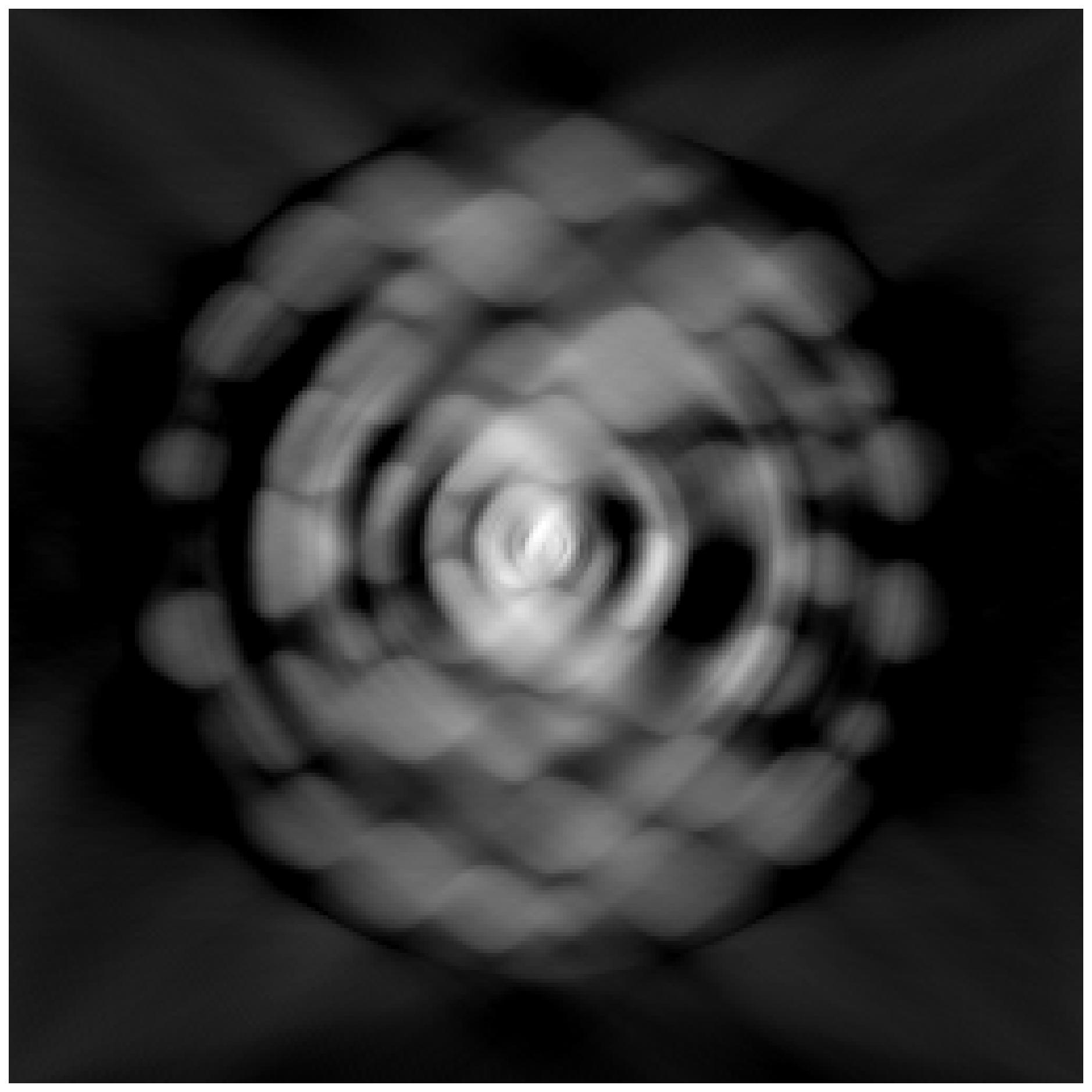}
\label{fig:reconstructions128_linearInterpolation}}
\hfill
\subfloat[Reconstruction with missing acquisitions replaced by CAD-expected ones]{\includegraphics[width=0.2\textwidth]{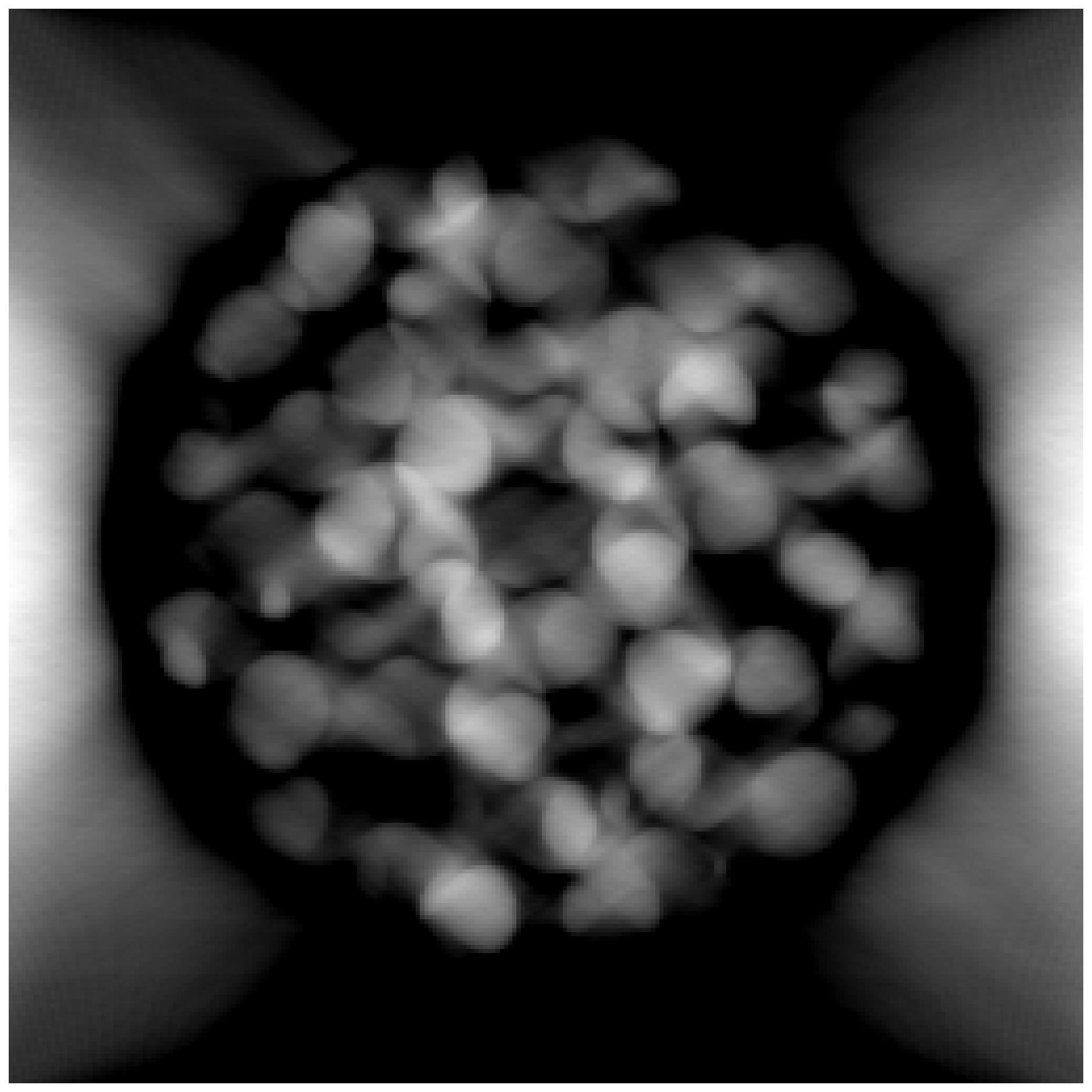}
\label{fig:reconstructions128_inferenceCad}}
\hfill
\subfloat[Reconstruction with acquisitions inferred from the Unet without the CAD prior]{\includegraphics[width=0.2\textwidth]{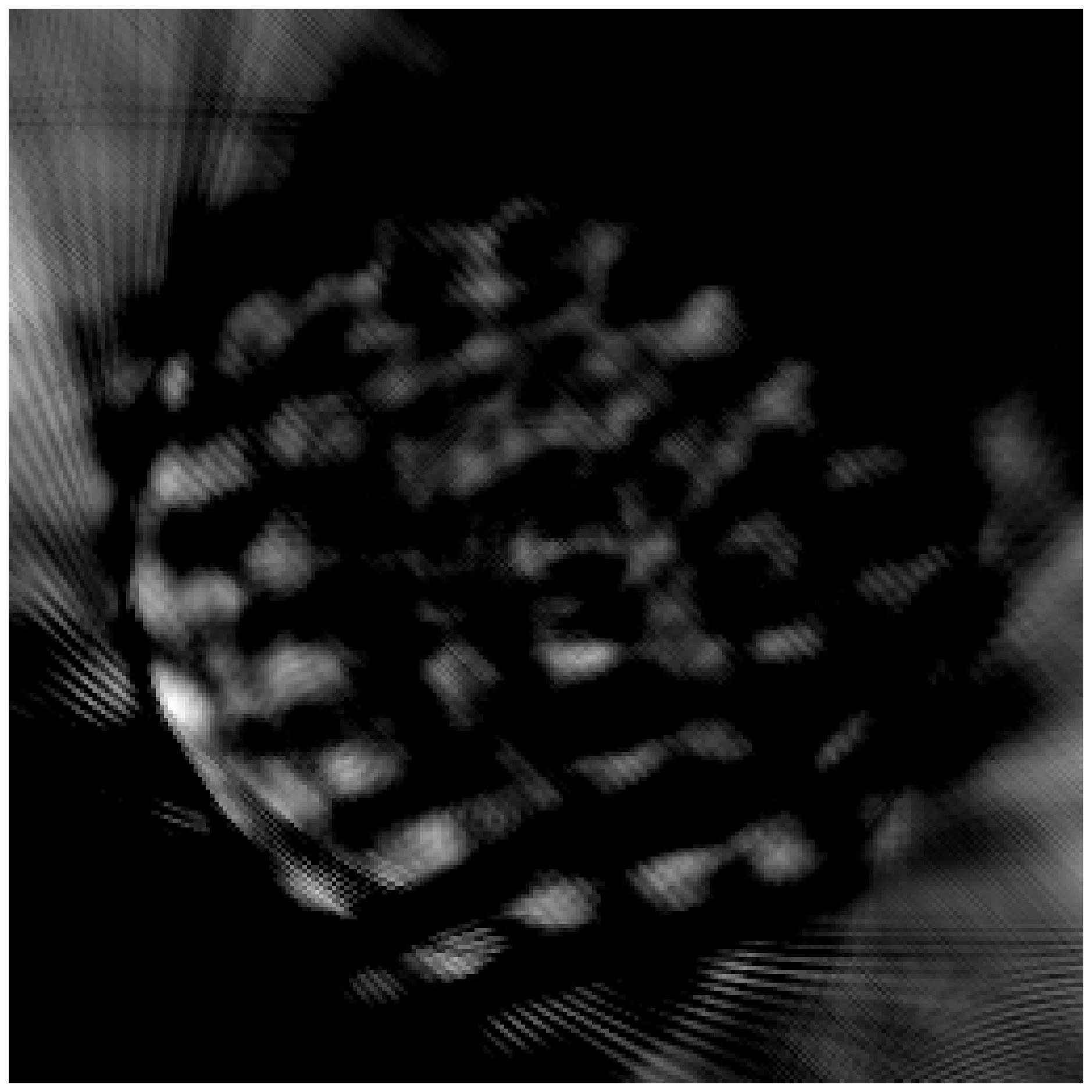}
\label{fig:reconstructions128_NoCadUnet}}
\hfill
\subfloat[Reconstruction with acquisitions inferred from the Unet with the CAD prior]{\includegraphics[width=0.2\textwidth]{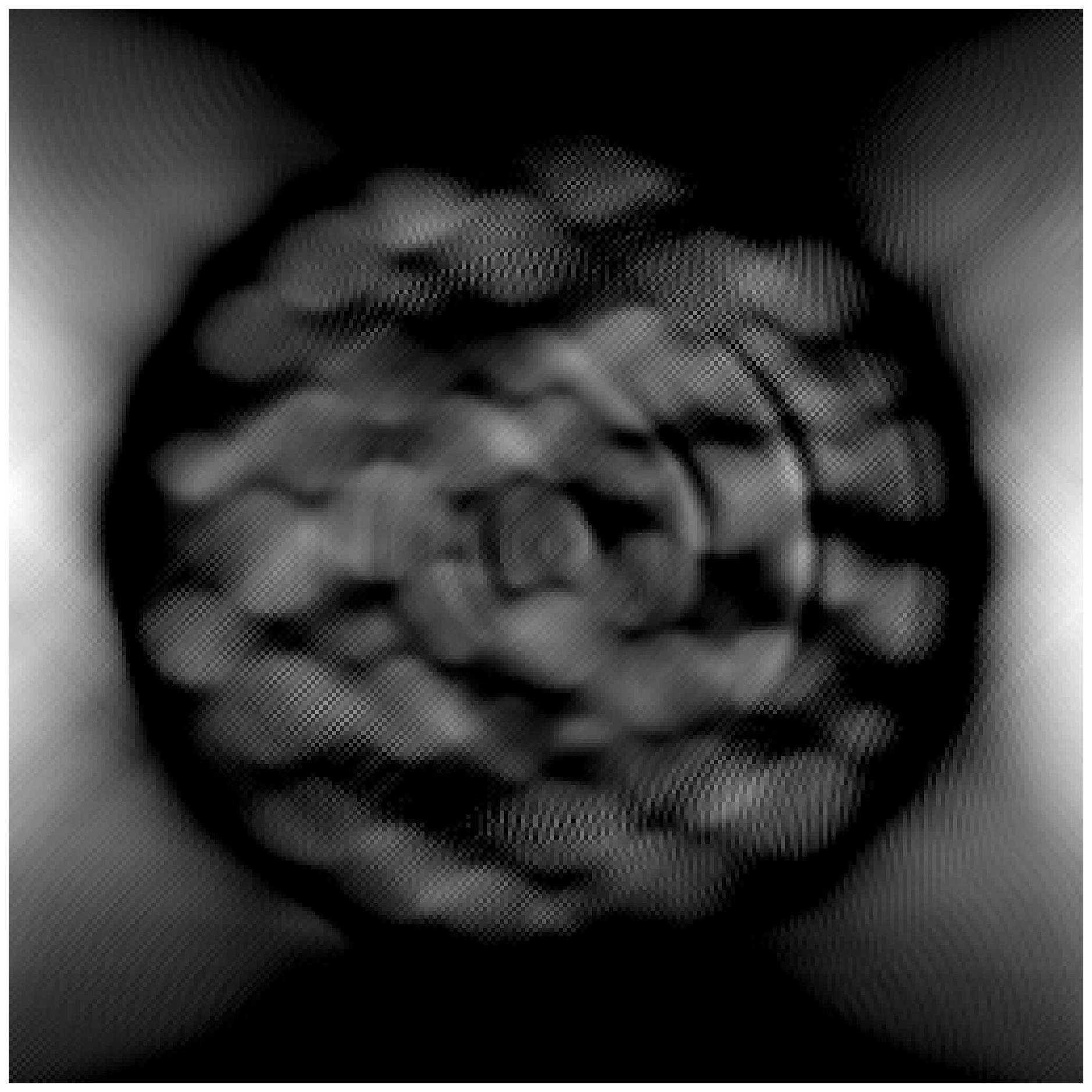}
\label{fig:reconstructions128_CadUnet}}
\hfill
\subfloat[Reconstruction with sinogram inpainted by the GAN.]{\includegraphics[width=0.2\textwidth]{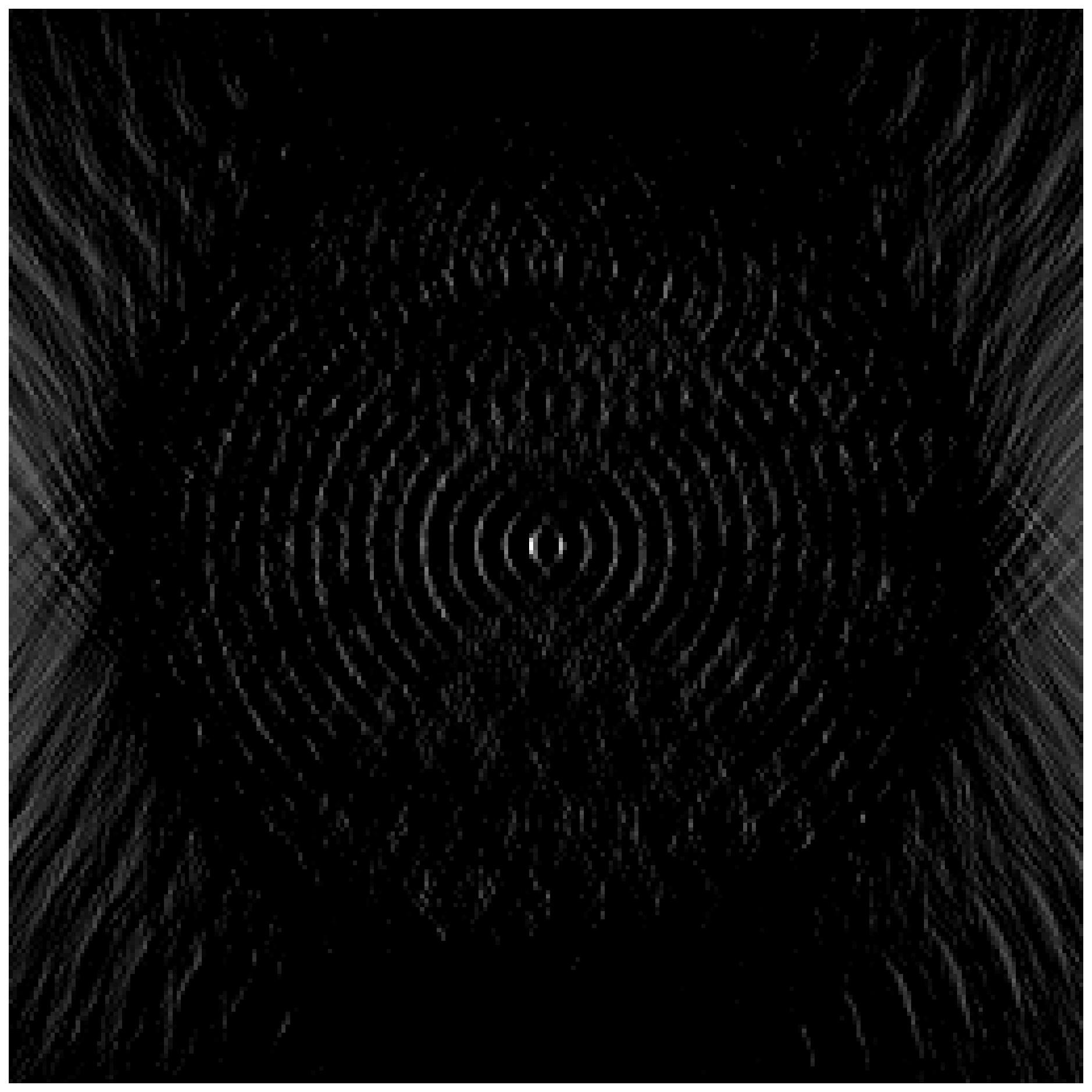}
\label{fig:reconstructions128_NoCadGan}}
\hfill
\subfloat[Reconstruction with missing acquisitions inferred from the pix2pix architecture using CAD prior.]{\includegraphics[width=0.2\textwidth]{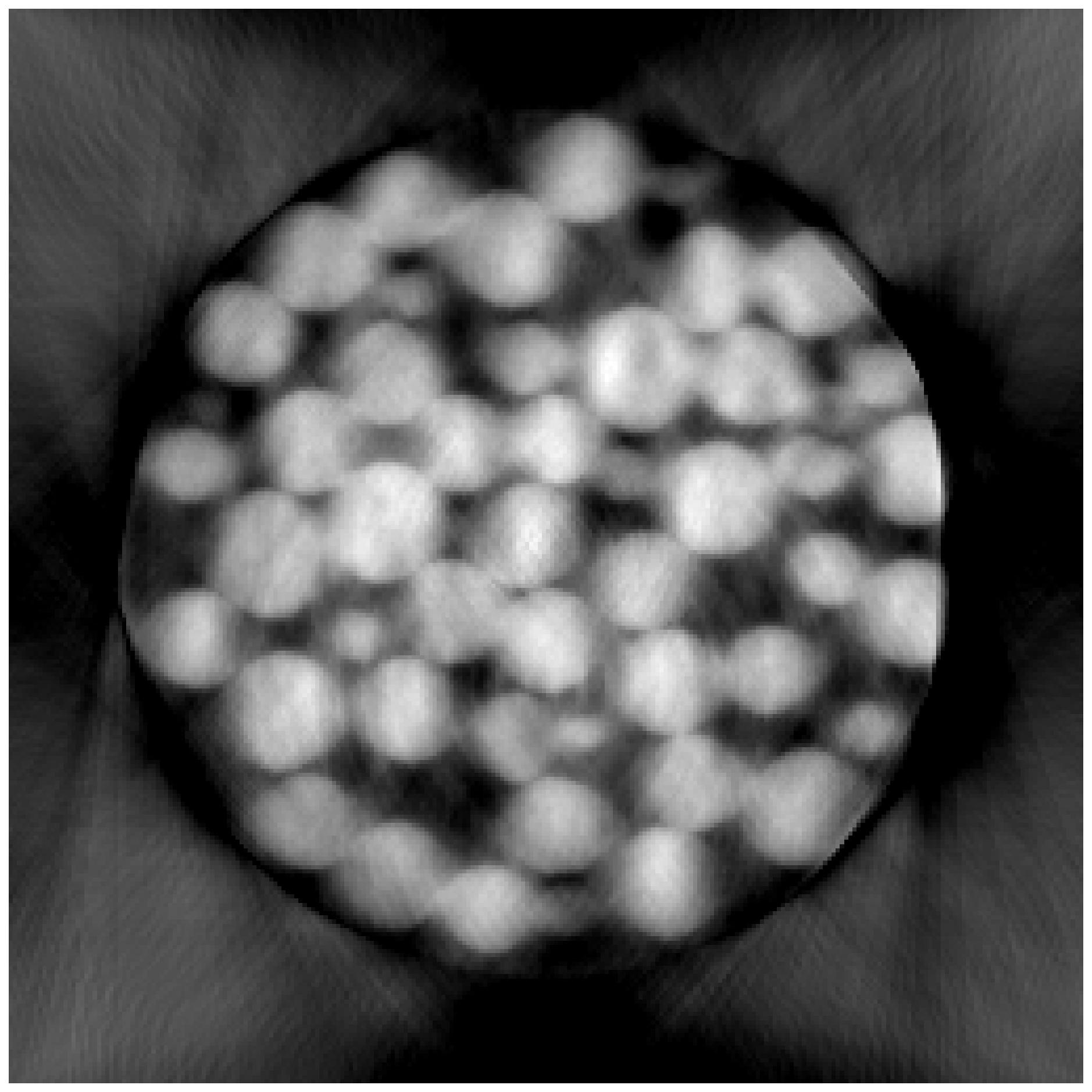}
\label{fig:reconstructions128_CadGan}}

\label{fig:reconstructions}
\caption{Reconstructions of the first sample of the SophiaBeads test dataset.(\ref{fig:reconstructions128_target}) is the target image, reconstructed from all 256 acquisitions and (\ref{fig:reconstructions128_noInterpolation}) is the image reconstructed from the scarce sinogram. (\ref{fig:reconstructions128_linearInterpolation}) to (\ref{fig:reconstructions128_CadGan}) are the images reconstructed from sinograms enhanced with various methods.}
\end{figure*}  

\begin{figure*}
\centering
\subfloat[Target reconstruction.]{\includegraphics[width=0.15\textwidth]{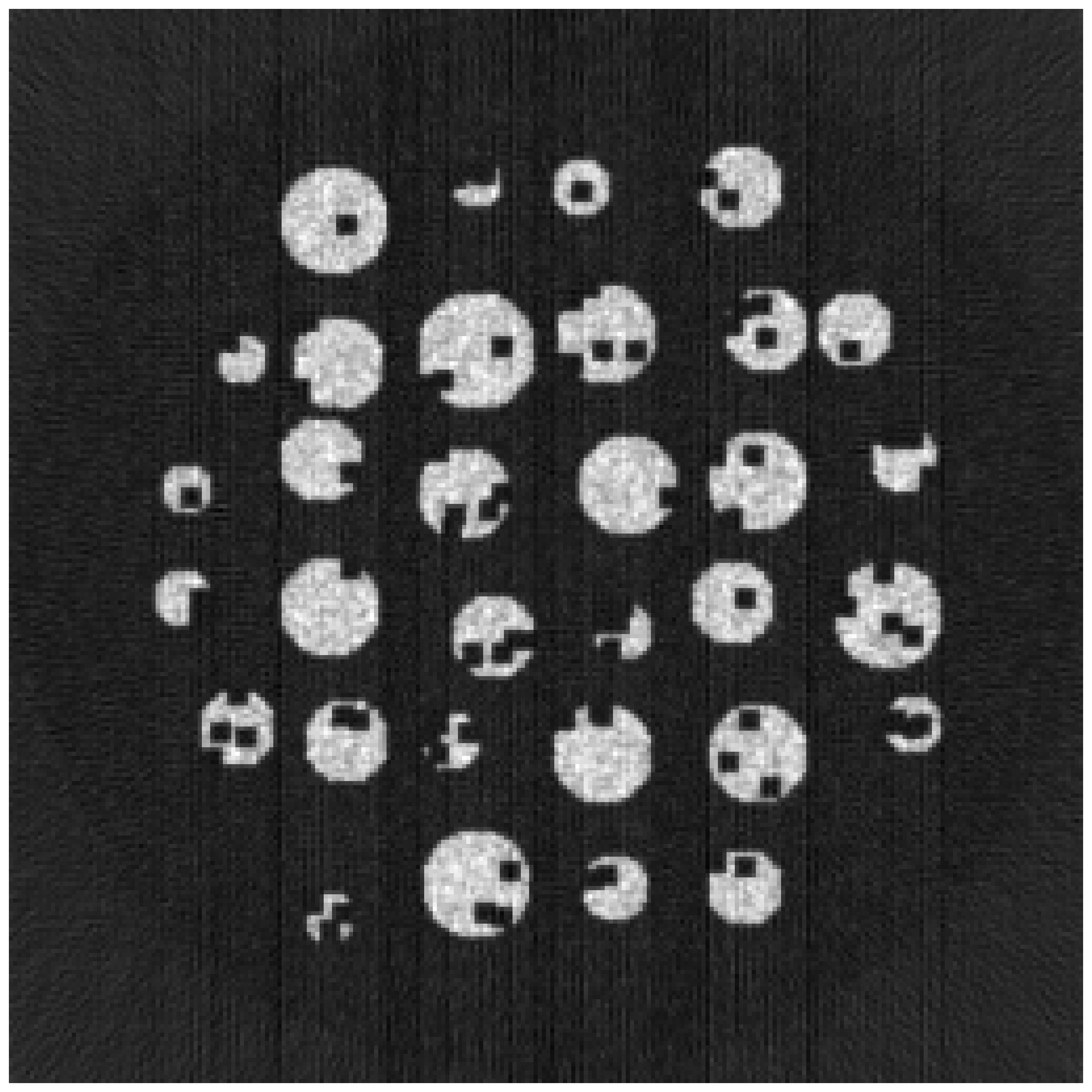}
\label{fig:holes_tar}}
\hfill
\subfloat[Reconstruction using our method.]{\includegraphics[width=0.15\textwidth]{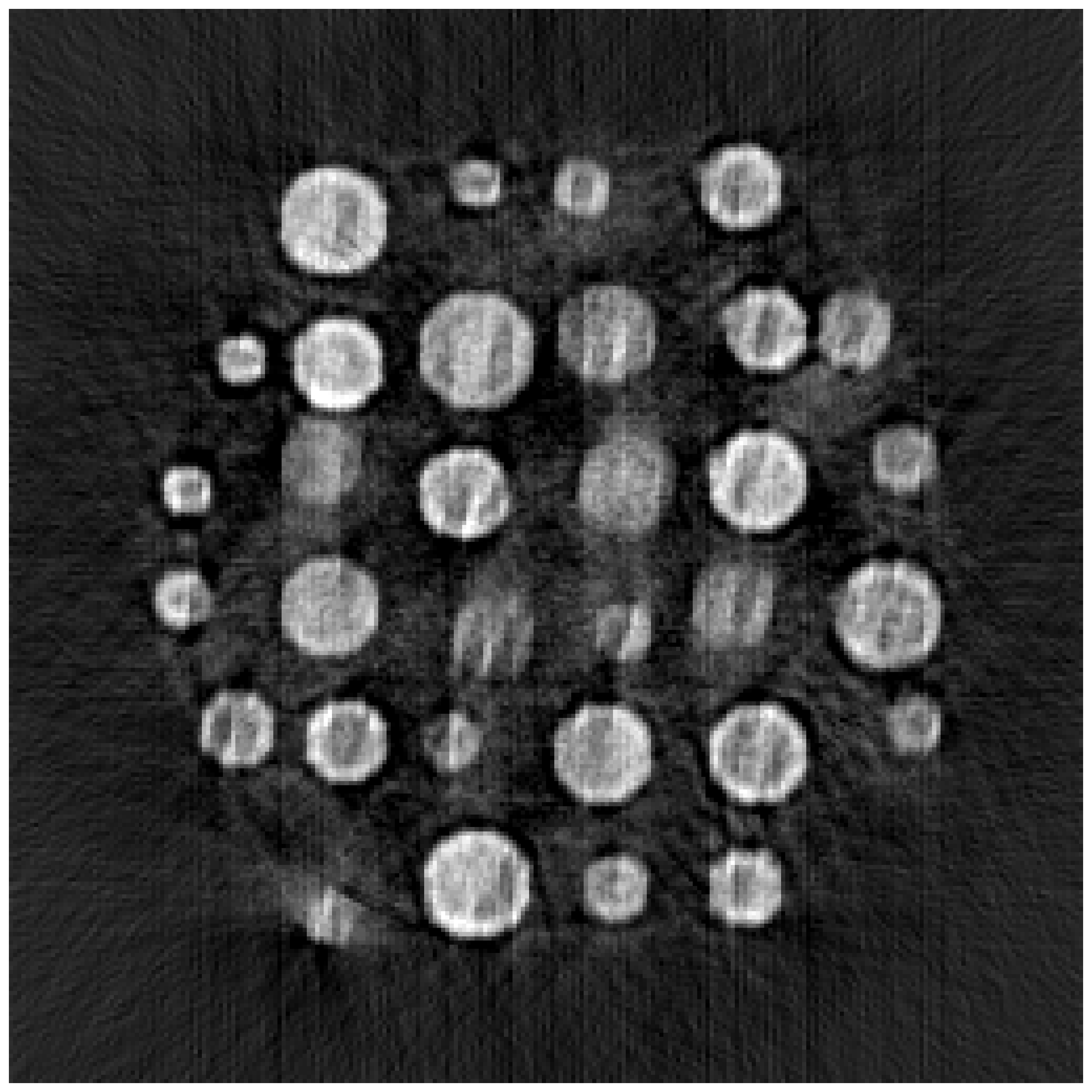}
\label{fig:holes_gan}}
\hfill
\subfloat[Difference between the image reconstructed with our method and the target reconstruction.]{\includegraphics[width=0.15\textwidth]{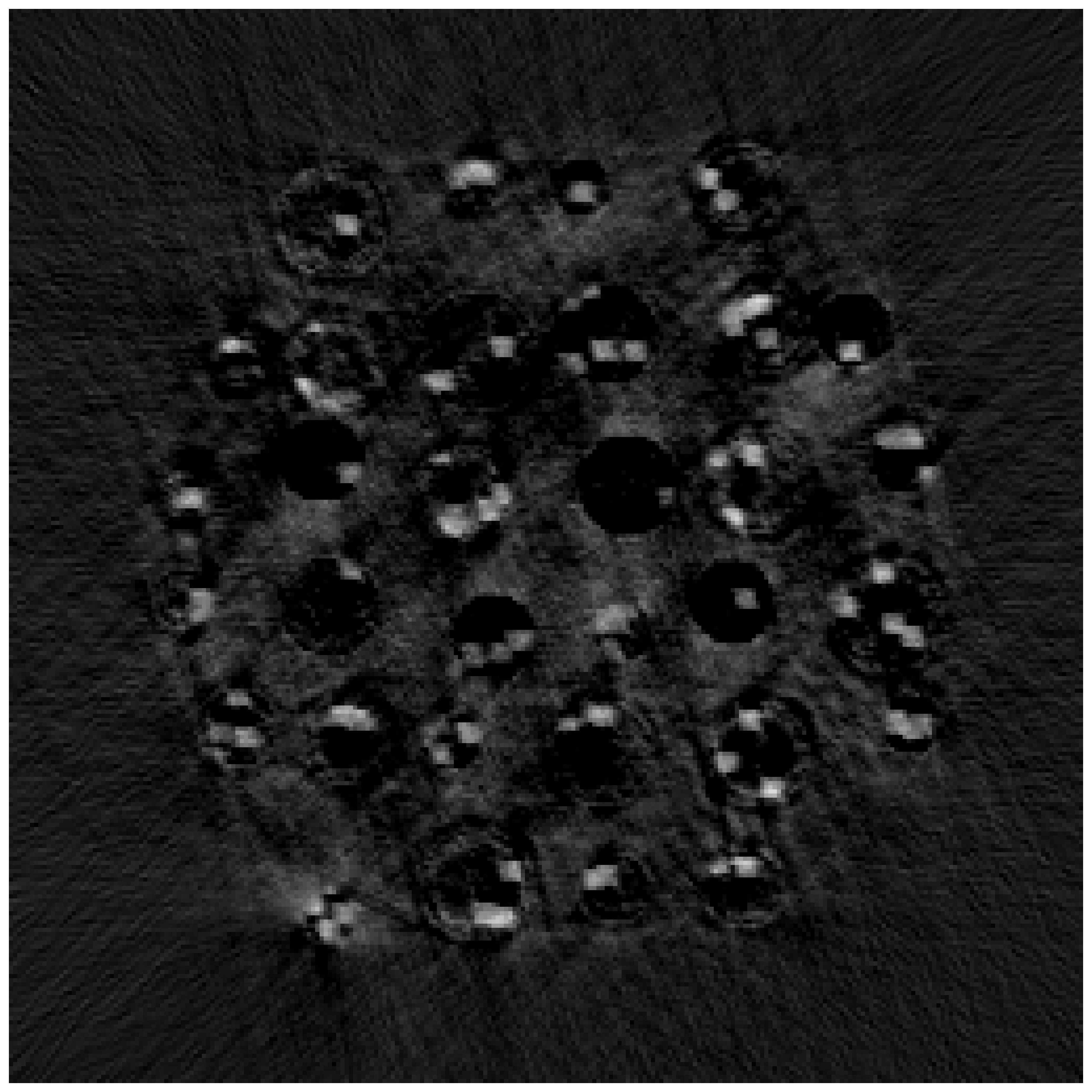}
\label{fig:holes_gan_tar}}
\hfill
\subfloat[Reconstruction using the shape prior.]{\includegraphics[width=0.15\textwidth]{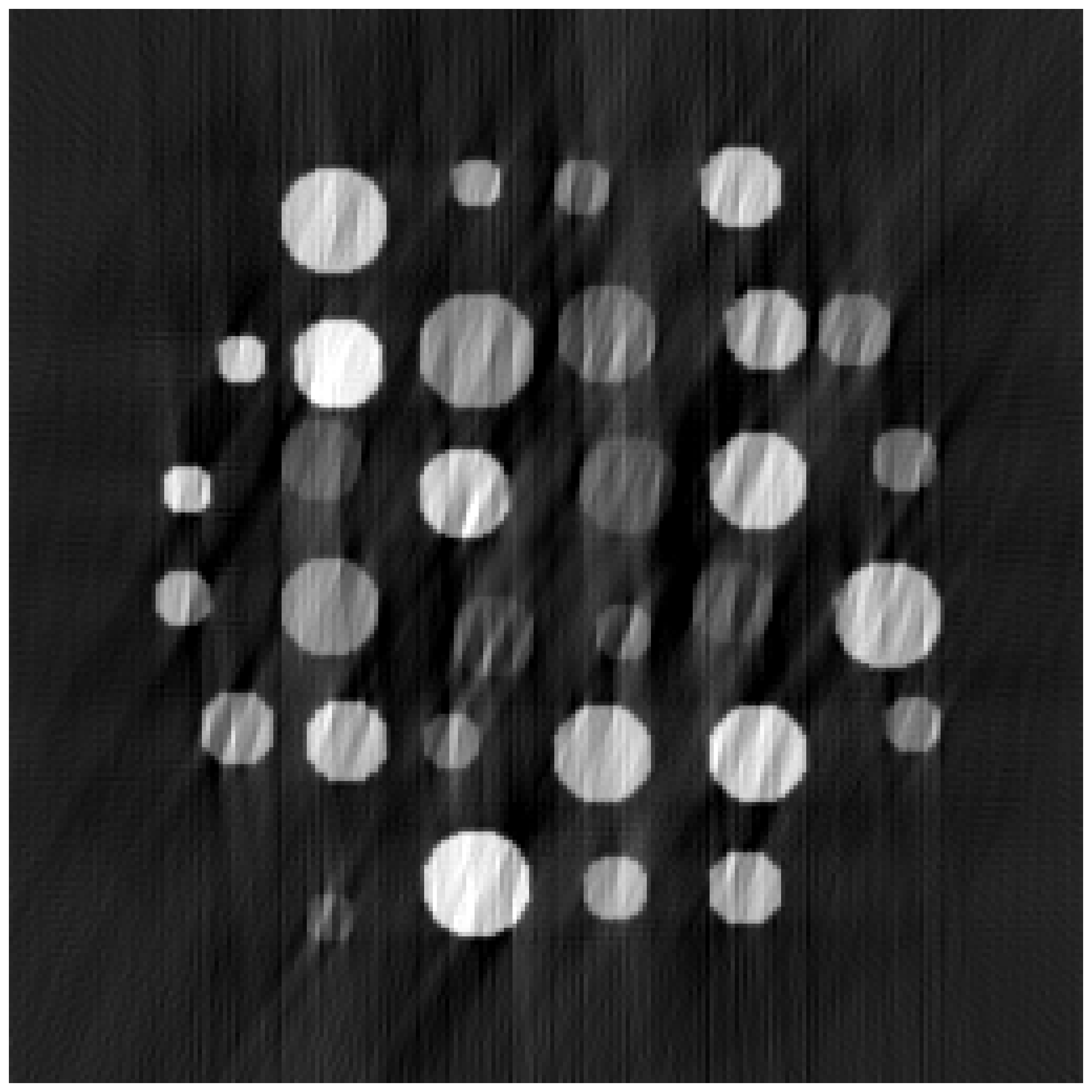}
\label{fig:holes_cad}}
\hfill
\subfloat[Difference between the image reconstructed with the shape prior and the target reconstruction.]{\includegraphics[width=0.15\textwidth]{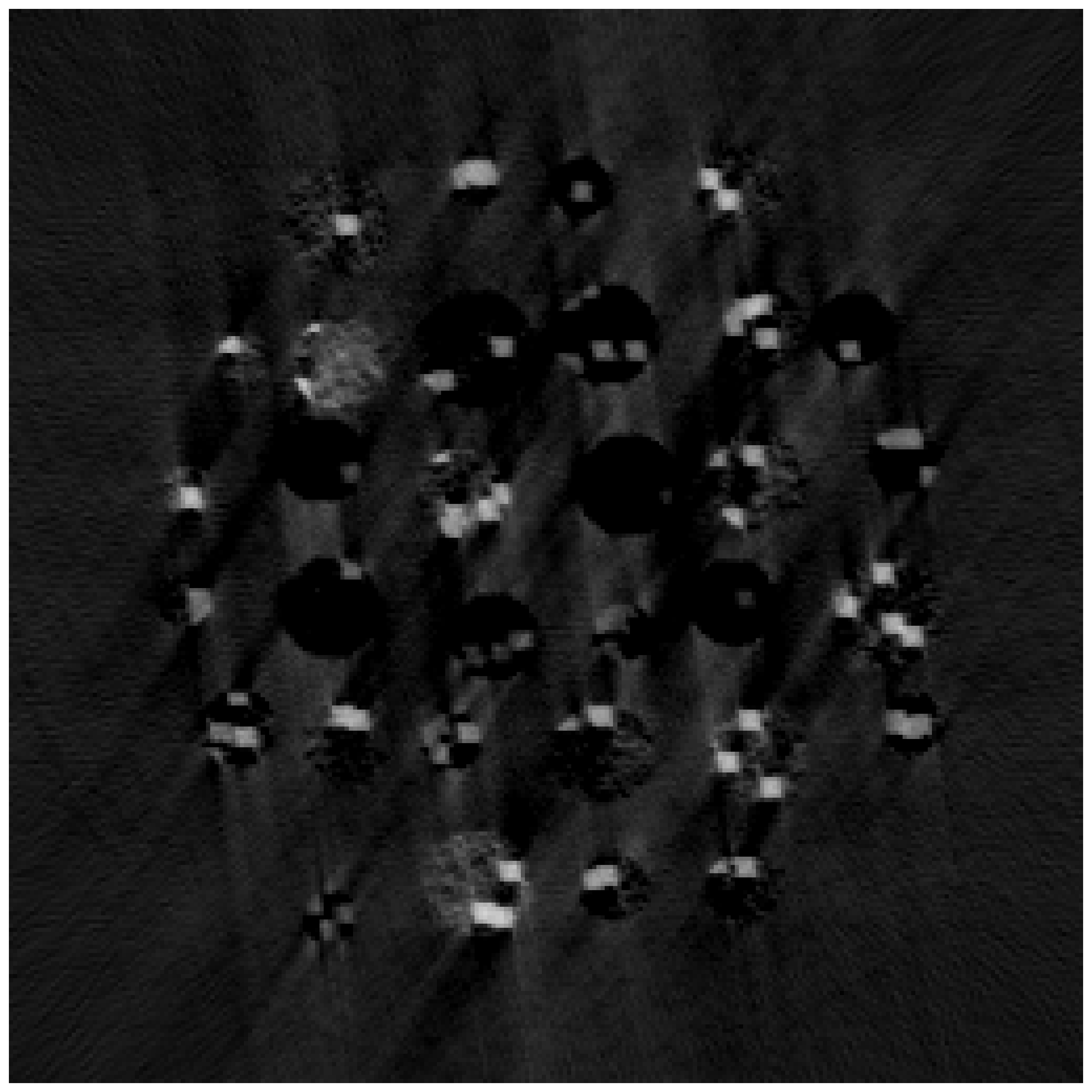}
\label{fig:holes_cad_tar}}
\caption{Understanding the action of the GAN on objects with internal defects not expected by the shape prior. (\ref{fig:holes_tar}) shows the target reconstruction, (\ref{fig:holes_gan}) shows the reconstruction with our method and (\ref{fig:holes_gan_tar}) shows the difference between (\ref{fig:holes_gan}) and (\ref{fig:holes_tar}). (\ref{fig:holes_cad}) shows the reconstruction with the shape prior and (\ref{fig:holes_cad_tar}) shows the difference between (\ref{fig:holes_cad}) and (\ref{fig:holes_tar}).}
\label{fig:images_with_holes}
        
\end{figure*}